\documentclass[12pt,reqno]{article}%{smfart}
\usepackage{times}%\usepackage{mathptmx}
\usepackage{amsmath}
\usepackage{amsthm}
\usepackage{amsfonts}
\usepackage{amssymb}
\usepackage{epic}
\usepackage{eepic}
\usepackage{times, macros,color}
\usepackage{graphics}
\usepackage[matrix,arrow]{xy}

\theoremstyle{plain}

\theoremstyle{remark}

\newcommand{\sst}{\scriptscriptstyle}

\newcommand{\pa}{\partial}
\newcommand{\ot}{\otimes}
\newcommand{\crd}{}
\newcommand{\ra}{\to}

\newcommand{\fsl}{{\mathfrak s}{\mathfrak l}}

\newcommand{\on}{\operatorname}

\newcommand{\C}{{\mathbb C}}

\newcommand{\ga}{\gamma}

\newcommand{\de}{\delta}
\newcommand{\De}{\Delta}
\newcommand{\ep}{\epsilon}
\newcommand{\la}{\lambda}

\newcommand{\si}{\sigma}

\newcommand{\vf}{\varphi}

\newcommand{\CB}{{\mathcal B}}
\newcommand{\CC}{{\mathcal C}}
\newcommand{\CD}{{\mathcal D}}
\newcommand{\CE}{{\mathcal E}}
\newcommand{\CF}{{\mathcal F}}

\newcommand{\CH}{{\mathcal H}}

\newcommand{\CL}{{\mathcal L}}
\newcommand{\CM}{{\mathcal M}}

\newcommand{\CO}{{\mathcal O}}  

\newcommand{\CS}{{\mathcal S}}
\newcommand{\CZ}{{\mathcal Z}}

\newcommand{\glos}{${}^{g)}$}
\newcommand{\hfg}{\hat{{\mathfrak g}}}

\newcommand{\BR}{{\mathbb R}}

\newcommand{\BC}{{\mathbb C}}

\newcommand{\BZ}{{\mathbb Z}}

\newcommand{\rf}[1]{(\ref{#1})}

\begin{document}\thispagestyle{empty}
\title{Supersymmetric field theories and geometric Langlands: The other side of the coin}
\author{Aswin Balasubramanian, J\"org Teschner}
\address{Department Mathematik, \\Universit\"at Hamburg, \\Bundesstraße 55, 
\\20146 Hamburg, Germany\\[1ex]and:\\[1ex]
DESY theory, \\
Notkestrasse 85, \\
22607 Hamburg, Germany}
\maketitle
\begin{center}{\bf Abstract}\end{center}
\begin{quote}
{\small
This note announces results on the relations between the approach of Beilinson and
Drinfeld  to the geometric Langlands correspondence based on conformal field theory,
the approach of Kapustin and Witten based on $N=4$ SYM, and the AGT-correspondence.
The geometric Langlands correspondence is described as the Nekrasov-Shatashvili limit 
of a generalisation of the AGT-correspondence in the presence of surface operators. 
Following the approaches of Kapustin - Witten and Nekrasov - Witten we interpret some
aspects of the resulting picture using an effective description in terms of two-dimensional 
sigma models having Hitchin's moduli spaces as target-manifolds.
}
\end{quote}

\section{Introduction}

Some remarkable connections between supersymmetric gauge theories and conformal 
field theory (CFT) have been discovered in the last few years. One of the most explicit 
connections was discovered by Alday, Gaiotto and Tachikawa \cite{AGT}, nowadays often referred to 
as AGT-correspondence. It overlaps with another family of results for which 
Nekrasov and collaborators have introduced the name BPS/CFT-correspondence, 
see \cite{N15} for the first in a series of papers on this subject and references to previous work in this
direction.  An older development, the relations between the
$N=4$ super-Yang-Mills theory (SYM) and the geometric Langlands correspondence 
exhibited by Kapustin and Witten \cite{KW},  naturally fits into the emerging picture 
\cite{NW,NRS}. One may note, however, that 
the approach of Beilinson and Drinfeld to the geometric Langlands 
correspondence has close connections to CFT, see \cite{Fr07} for a review and further references, 
which are not obvious in the approach of Kapustin and Witten.
Some connections between these subjects
have been discussed in \cite{NW,T10,NRS,GW}, 
and the recent work \cite{KP,AFO} indicates that at least a part of these relations admit a 
further deformation, motivated by supersymmetric gauge and string theory.
However, the picture still seems to be incomplete in many respects.

This note, prepared for the proceedings of String-Math 2016, announces 
results shedding some light on the relations between the approaches to the
geometric Langlands correspondence of Beilinson and Drinfeld, Kapustin and Witten, 
and the AGT-correspondence. We will mostly
restrict attention to the cases where the underlying Lie-algebra is $\mathfrak{sl}_2$
in order to keep the length of this note within reasonable bounds.
The results described here are part of a larger project being pursued by the authors
together with Ioana Coman-Lohi.
A series of publications containing more details, an extended discussion of mathematical
aspects,  and 
a discussion of the higher rank cases is in preparation.

In the main text we will freely use several standard definitions and results concerning  
Hitchin's moduli spaces. A very brief summary is 
collected in Appendix \ref{Gloss} in the form of a glossary. 
If a glossary entry exists for a term, its first occurrence will appear with a superscript as in term$^{g)}$.

\section{Review}

\subsection{What is the geometric Langlands correspondence?}

The geometric Langlands correspondence is often loosely formulated as a correspondence which assigns
$\mathcal{D}$-modules on ${\rm Bun}_G$ to ${}^LG$-local systems\glos on a Riemann surface $C$. 
${}^LG$ is the Langlands dual group of a simple complex Lie group $G$.
The  ${}^LG$-local systems appearing in this correspondence can be represented by 
pairs $(\mathcal{E},\nabla')$ composed of a holomorphic ${}^LG$-bundle $\mathcal{E}$ 
with a holomorphic connection $\nabla'$, 
or equivalently by the representations $\rho$ of the fundamental group $\pi_1(C)$ defined from the holonomies of 
$(\mathcal{E},\nabla')$. We will mostly be interested in the 
case of {\it irreducible} ${}^LG$-local systems. The corresponding 
$\mathcal{D}$-modules on ${\rm Bun}_G$ can be described more concretely as systems of 
partial differential equations taking the form of eigenvalue equations 
$D_i f= E_if$ for a family of differential operators $D_i$ on ${\rm Bun}_G$ obtained by quantising 
the Hamiltonians of Hitchin's integrable system. A more ambitious version of the 
geometric Langslands correspondence has been formulated in \cite{AG} in which 
it becomes necessary to extend it to certain classes of reducible local systems.

Some of the original approaches to the geometric Langlands correspondence
start from the cases where the ${}^LG$-local systems are {\it opers}$^{g)}$, pairs $(\mathcal{E},\nabla')$
in which  $\nabla'$ is gauge-equivalent to a certain standard form. The space of opers 
forms a Lagrangian subspace in the moduli space of all local systems. 
The CFT-based approach of Beilinson and Drinfeld  constructs for each oper an 
object in the category of 
$\mathcal{D}$-modules on ${\rm Bun}_G$ as conformal blocks of the affine Lie algebra 
$\hat{\mathfrak g}_k$ at the critical level $k=-h^\vee$. The Ward-identities characterising the conformal blocks 
equip the sheaves of conformal blocks with a $\mathcal{D}$-module structure.
The universal enveloping algebra
$\mathcal{U}(\hat{\mathfrak g}_{k})$
has a large center at $k=-h^\vee$, isomorphic to the space of  ${}^L\mathfrak{g}$-opers 
on the formal disc \cite{FF92}.  This can be used to show that the $\mathcal{D}$-module structure 
coming from the Ward identities can be described by the system of eigenvalue equations 
$D_i f= E_if$ for the quantised Hitchin Hamiltonians, with eigenvalues $E_i$ parameterising the 
choice of opers \cite{Fr07}.

There exists an extension of the Beilinson-Drinfeld construction of the geometric Langlands correspondence 
described in \cite[Section 9.6]{Fr07}
from the case of opers to general irreducible local systems. 
It is based on the fact that 
such local systems are always gauge-equivalent to opers with certain extra singularities 
\cite{Ar16}. The construction of Beilinson and Drinfeld associates to such opers 
conformal blocks of $\hat{\mathfrak g}_{-h^{\vee}}$ with certain degenerate 
representations induced from the finite-dimensional representations of $\mathfrak{g}$ inserted 
at the extra singularities. We may in this sense regard the geometric Langlands 
correspondence for general irreducible local
systems as an extension of the correspondence that exists for ordinary, non-singular opers.  
This  point of view will turn out to be natural from the perspective we will propose in this paper.
Let us also remark that the construction of Beilinson-Drinfeld plays an important role in the 
outline given in \cite{Ga15} for a proof of the strengthened geometric Langlands conjecture formulated in \cite{AG}.

For future reference let us note that the Beilinson-Drinfeld construction describes the 
$\mathcal{D}$-modules appearing in the geometric Langlands correspondence as 
spaces of conformal blocks which, naturally being fibered over ${\rm Bun}_G$ on the one hand,  
are furthermore getting fibered over the spaces of irreducible ${}^LG$-local systems 
on the other hand.

%(Add sthg. on $\mathcal{D}$-modules holonomic, flat connection on subspace of ``very stable'' bundles?)

%The construction of Beilinson-Drinfeld provides a functor from the category 
%${\rm Coh(Loc_{\check{\fg}}(C))}$ 
%of coherent sheaves on the moduli space of local systems on $C$ to the category 
%$\mathcal{D}\text{-}{\rm mod}({\rm Bun}_G)$ of
%D-modules on ${\rm Bun}_{G}$. This functor takes the skyscraper sheaf $\mathcal{O}_{\si}$ associated to 
%a local system $\si$ to a D-module $\Delta_\si$ on ${\rm Bun}_G$. The fibers  $\Delta_{\si|x}$ of the  D-module
%$\Delta_\si$ over a family of points $\{\mathcal{B}_x,x\in\mathcal{X}\}$,  
%$\mathcal{B}_x\in{\rm Bun}_G$, are isomorphic to the space 
%${\rm CB}(C,\mathcal{B}_x,{\mathbb M}_{\rho(\si)})$ of conformal blocks on $C$  associated to the collection of 
%representations ${\mathbb M}_{\rho}=\{\mathcal{M}_{\rho_y},y\in C\}$ of the current algebra 
%$\hat{\fg}_{-h^\vee}$ 
%at the critical level.  $\rho(\si)$ is the oper associated to the local system $\si$ via the Riemann-Hilbert 
%correspondence.
%The representations $\mathcal{M}_{\rho_y}$ are parameterised by the values of the central 
%elements of $\mathcal{U}(\hat{\fg}_{-h^\vee})$ specified by the restriction $\rho_y$ of the oper $\rho$ to a 
%formal disc around $y\in C$.

\subsection{Geometric Langlands correspondence - approach of Kapustin-Witten}

The variant of the geometric Langlands correspondence proposed by Kapustin and Witten \cite{KW} 
is based on the consideration of $N=4$ SYM theory with gauge group {$G_c$, a compact real form of $G$},
on four-manifolds of the form 
$\Sigma\times C$, 
where $C$ is a Riemann surface. 
Compactification on $C$ allows one to represent
the topologically twisted version of $N=4$ SYM effectively 
by a topologically twisted $2d$ sigma-model  with target being the Hitchin moduli space$^{(g)}$ 
$M_H(G)$ on $\Sigma$. The complete
integrability of the Hitchin moduli space, as is manifest in the description of 
$M_H(G)$ as a torus fibration, allows one to describe the consequences of the 
S-duality of $N=4$ SYM theory as the SYZ mirror symmetry relating 
the $2d$ sigma-models with target $M_H(G)$  and $M_H({}^L G)$, respectively. 

In order to relate this to the geometric Langlands correspondence,
Kapustin and Witten consider the cases when  $\Sigma=R\times I$, $I=[0,\pi]$. Supersymmetric 
boundary condition of $N=4$ SYM theory will upon compactification on $C$ define
boundary conditions of the 
$\mathcal{N}=(4,4)$ sigma model with target $\CM_H(G)$ on $\Sigma$.
Let  $\mathfrak{B}$ be the category having as objects
boundary conditions $\mathsf{B}$  called branes 
preserving the maximal number of supersymmetries, with 
morphisms being the  spaces of "open string" states ${\rm Hom}_{\CM_H(G)}^{}(\mathsf{B}_1,\mathsf{B}_2)$
of the sigma model on the strip $\BR\times I$, having  associated boundary conditions $\mathsf{B}_1$ and 
$\mathsf{B}_2$ to the boundaries $\BR\times \{0\}$ and $\BR\times \{\pi\}$, respectively.
%\end{itemize}

A distinguished role is played by the so-called canonical coisotropic brane $\mathsf{B}_{\rm cc}$ \cite{KW,NW}. 
The vector space  $\mathcal{A}_{\rm cc}
={\rm Hom}_{\CM_H(G)}^{}(\mathsf{B}_{\rm cc},\mathsf{B}_{\rm cc})$ has a natural algebra structure with product
defined by "joining open strings". %This algebra will be called $\mathcal{A}_\hbar(G)$, where $\hbar=\ep_1/\ep_2$.
The spaces $\CH(\mathsf{B})
={\rm Hom}_{\CM_H(G)}^{}(\mathsf{B}_{\rm cc},\mathsf{B})$ are left modules
over the algebra $\mathcal{A}_{\rm cc}$ with action defined by "joining open strings" from 
$\mathcal{A}_{\rm cc}$ on the left boundary of the strip $I$. Kapustin and Witten
argue that the algebra $\mathcal{A}_{\rm cc}$ contains the algebra of global differential operators on 
${\rm Bun}_G$.  It follows that the spaces $\CH(\mathsf{B})$  represent
$\mathcal{D}$-modules on ${\rm Bun}_G$. 

The reduction of Wilson- and 't Hooft line operators with support on $\BR\times\{x\}\times P$, $x\in I$, 
$P\in C$, to the two dimensional TQFT defines
natural functors on the category of branes, inducing modifications of the spaces $\CH(\mathsf{B})$. 
The functors defined in this way are identified in \cite{KW} with the Hecke functors in the geometric
Langlands correspondence. 
For some branes $\mathsf{B}$ one may represent for each fixed $P\in C$ 
the resulting modification 
as the tensor product of $\CH(\mathsf{B})$ with a finite-dimensional 
representation $V$ of ${}^LG$.  One says that the brane $\mathsf{B}$ 
satisfies the Hecke eigenvalue property if the family of modifications obtained by varying
the point $P\in C$ glues into a local system.

A family of branes $\mathsf{F}_\mu$ is identified in \cite{KW} having this property.
The branes $\mathsf{F}_\mu$ are supported on fibers of the Hitchin's torus fibration.
S-duality of $N=4$ SYM gets represented within 
the sigma model with target $\CM_H(G)$ as a variant of 
SYZ mirror symmetry, relating the branes $\mathsf{F}_\mu$ to 
branes the dual sigma model with target $\CM_H({}^LG)$ represented by
skyscraper sheaves $\check{\mathsf{F}}_\mu$ 
having pointlike support at $\mu\in\CM_H({}^LG)$.

\subsection{AGT-correspondence - approach of Nekrasov-Witten}

Alday, Gaiotto and Tachikawa discovered a relation between the instanton partition 
functions of certain $N=2$ supersymmetric gauge theories and conformal blocks
of the Virasoro algebra \cite{AGT}. This discovery has stimulated a lot of work leading in
particular to  various generalisations of such relations. In an attempt to 
explain the relations discovered in \cite{AGT}, Nekrasov and Witten considered
four-dimensional $N=2$ supersymmetric gauge theories of class $\mathcal{S}$ obtained from
the maximally supersymmetric six-dimensional QFT
on manifolds of the form
$\CM^4\times C$ by compactification on the Riemann surface $C$. For the case associated to the Lie
algebra $\mathfrak{g}=\mathfrak{sl}_2$ one has 
weakly coupled Lagrangian descriptions of the resulting theory associated to the choice of 
a pants decomposition $\si$ of $C$.
For four-manifolds 
$\CM^4$ which can be represented as a fibered product locally of the form $\mathbb{R}\times I\times S^1\times
S^1$ 
it is argued in \cite{NW} that (i) an $\Omega$-deformation with parameters $\ep_1,\ep_2$ can be defined, and 
(ii) an effective representation is obtained by compactification on $ S^1\times
S^1$ in terms of a
sigma-model  with target $\CM_H(G)$ on $R\times I$.
The coupling parameter of this sigma model is $\ep_1/\ep_2$. 

The end points of the interval 
$I$ in the representation $\CM^4\simeq\mathbb{R}\times I\times S^1\times
S^1$ correspond to points where $\CM^4$ is perfectly regular. One
must therefore have distinguished boundary condition in the 
sigma-model  with target $\CM_H(G)$ on $R\times I$ describing the compactification 
of a class $\mathcal{S}$ theory on $\CM^4$. When the compactification yields
a sigma model with target $\CM_H(G)$, it is argued in \cite{NW} that the corresponding 
boundary conditions are described by a variant  $\mathsf{B}_{\rm cc}$ of the 
canonical coisotropic brane at $\BR\times\{0\}$, and a new type of brane  called the 
``brane of opers'', here denoted by $\mathsf{B}_{\rm op}$, respectively.\footnote{The branes denoted
$\mathsf{B}_{\rm cc}$ in this context are similar but not identical with the brane considered in \cite{KW}.
The paper 
\cite{NW} used the notation $B_{N'}$ for the brane denoted $\mathsf{B}_{\rm op}$ here}
The brane $\mathsf{B}_{\rm op}$ is the mirror dual of $\mathsf{B}_{\rm cc}$,
and it is proposed in \cite{NW} that
the brane $\mathsf{B}_{\rm op}$ is a Lagrangian %$(A,A,B)$-
brane supported on the 
variety of opers within $M_H(G)$.  
 
In \cite{NW} it is furthermore proposed that the space 
$\CH={\rm Hom}_{\CM_H(G)}^{}(\mathsf{B}_{\rm cc},\mathsf{B}_{\rm op})$ 
can be identified
with the space of Virasoro conformal blocks. In order to motivate this identification,
Nekrasov and Witten note that the algebra $\mathcal{A}_{\rm cc}^{\hbar}(G)=
{\rm Hom}_{\CM_H(G)}^{}(\mathsf{B}_{\rm cc},\mathsf{B}_{\rm cc})$ with $\hbar=\ep_1/\ep_2$
is isomorphic to
the algebra of Verlinde line operators acting on 
the space of Virasoro conformal blocks. 
Mirror symmetry produces a dual description of 
$\CH(G)\simeq\CH({}^{\rm L}G)$ as  the space
${\rm Hom}_{\CM_H({}^LG)}^{}({\mathsf{B}}_{\rm op}',{\mathsf{B}}_{\rm cc}')$,
with ${\mathsf{B}}_{\rm op}'$ and ${\mathsf{B}}_{\rm cc}'$ being 
close relatives of ${\mathsf{B}}_{\rm op}$ and ${\mathsf{B}}_{\rm cc}$, respectively,  with modified 
SUSY invariance properties.
In the dual representation one has an obvious right action of the algebra 
$\check{\mathcal{A}}_{\rm cc}^{1/\hbar}({}^{L}G)
={\rm Hom}_{\CM_H({}^LG)}^{}(\mathsf{B}_{\rm cc}',\mathsf{B}_{\rm cc}')$ 
with action defined by "joining open strings" on the right boundary of the strip $I$. 
The existence of (almost) commuting actions of $\mathcal{A}_{\rm cc}^{\hbar}(G)$ and
$\check{\mathcal{A}}_{\rm cc}^{1/\hbar}({}^{L}G)$ is a characteristic feature of the space of Virasoro 
conformal blocks. 
 
\subsection{The other way around}

It is no accident that the work of Nekrasov and Witten \cite{NW} has many elements in common with 
the approach Kapustin and Witten \cite{KW}. A common root can be found in the fact that 
both the class $\CS$-theories  and $N=4$ SYM \cite{Wi09} can be obtained as compactifications
of the six-dimensional $(2,0)$-theory on six-manifolds $\CM^6=\CM^4\times C$, where
$C$ is a Riemann surface, and $\CM^4$ is a four-manifold locally represented
as a circle fibration locally of the form $\mathbb{R}\times I \times S^1\times S^1$.
Compactification on $C$ yields class $\CS$-theories \cite{GMN2}, 
while compactification on $S^1\times S^1$
yields $N=4$ SYM on $\mathbb{R}\times I \times C$, the set-up
considered in \cite{KW} as was further discussed in \cite{Wi09}. 

One should note, however, 
that different topological twists are used in the two compactifications, making the comparison
of the results somewhat subtle.
This fact can nevertheless be used to relate supersymmetric boundary conditions in
the 2d sigma model with target $\CM_H$ arising from compactification of class $\CS$-theories
to boundary conditions in $N=4$ SYM on $C$. These boundary conditions have been classified in 
the work of Gaiotto and Witten \cite{GW1}. 
In this way, one may establish a relationship between the canonical coisotropic brane 
and pure Neumann boundary conditions in $N=4$ SYM \cite{KW,GW}. 
Exchanging the  two circles in $S^1\times S^1$
gets related to the S-duality of $N=4$ SYM which implies relations 
between its boundary conditions studied in \cite{GW2}. This led \cite{GW} to relate the 
brane $\mathsf{B}_{\rm op}$, the mirror dual of
the canonical coisotropic brane in \cite{NW}, to the boundary condition descending from the so-called
Nahm pole boundary conditions in $N=4$ SYM, as will briefly be discussed in Subsection \ref{S:Nahm} below.

\subsection{Other approaches} 

As noted above, one needs to use different twists in the two reductions from six to four 
dimensions considered above. In order to get the set-up studied in \cite{NW}, for 
example, one needs to twist the $(2,0)$ theory on $\CM^4\times C$ in such a way that
it becomes  topological on $\CM^4$. A different twist is obviously needed to get 
the topologically twisted $N=4$ SYM on $\BR\times I\times C$ studied in \cite{KW} from 
the six-dimensional $(2,0)$ theory. 

In order to describe the
dimensional reduction of topologically twisted $N=4$ SYM on $\BR\times I\times C$  
one may find it natural to consider boundary conditions that are purely topological,
not depending on the complex structure on $C$. This point of view motivated Ben-Zvi
and Nadler \cite{BN16} to propose the Betti geometric Langlands conjecture as a 
purely topological variant of the 
geometric Langlands correspondence formulated in \cite{AG} that is
naturally adapted to the four-dimensional TQFT's studied in \cite{KW}.

%\subsection{Approach of Cordova and Jafferis -- \crd really needed?} 

Yet another approach towards understanding the AGT-correspondence was proposed
by Cordova and Jafferis in \cite{CJ}.
It starts from six-manifolds of the form 
$\CM^4\times C$, with $\CM^4$ being a squashed four-sphere, represented 
as three-sphere fibration over an interval $I_0$, locally $\CM^4\sim I_0 \times S^3_{\ep_1\ep_2}$ with 
$\ep_1,\ep_2$ now being the squashing parameters.
Considering partial topological twisting on $C$, it is argued that the consequence
of Weyl invariance of the $(2,0)$ theory with the twist on $C$ is the fact
that partition functions depend neither 
on the overall radius of $\CM^4$ nor on the volume of $C$. The effective description 
at small volume of $C$ will be given by the partition functions of class $S$ theories 
on (squashed) four-spheres studied in the work of Pestun \cite{Pe} and many subsequent publications. 
It is argued that an equivalent description in terms of a CFT on $C$ is obtained by 
first compactifying on the three spheres $S^3_{\ep_1\ep_2}$ appearing in the 
representation $\CM^4\sim I_0 \times S^3_{\ep_1\ep_2}$. Using five-dimensional SYM theory
as an intermediate step, Cordova and Jafferis found an effective description in terms of 
complex Chern-Simons theory on $I_0\times C$. The boundary conditions at the ends of $C$ are
determined by the Nahm pole boundary conditions studied in \cite{GW1,GW2,GW}. 
A generalised version of the usual  correspondence between
Chern-Simons theory and two-dimensional 
conformal field theory can then be invoked, with Nahm pole boundary conditions 
translating into the constraints reducing the WZNW models to the conformal 
Toda theories.

%\subsection{AGT-correspondence - approach of Cordova-Jafferis}

%For QFT-interpretation (III) consider compactification: 
%$(2,0)$-theory on $M^6\simeq \BR\times S^3 \times C$ \\$\longrightarrow$ 
%complex Chern-Simons theory on $\BR\times C$ \cite{CJ}.

%

\section{Defects of co-dimension two and surface operators}\label{codim2}

M theory suggests a natural extension of the set-up considered above. Viewing the 
six-dimensional $(2,0)$ theory as the effective theory on a stack of M5-branes in M-theory 
leads to
a natural extension of the set-up by an additional stack of M5'-branes sharing part of the support
with the original M5-branes, as indicated in the following table:
\begin{center}
\begin{tabular}{l || c|c|c|c|c|c|c|c|c|c|c}
Brane & 0 & 1 & 2 & 3 & 4 & 5 & 6 & 7 & 8 & 9 & 10 \\ \hline\hline
M5  & x & x & x & x &   &   & x &   &   &   & x  \\
M5' & x & x &   &   &   &   & x &   & x & x & x \\
\end{tabular}
\end{center}
Depending on the type of compactification considered one gets surface operators in class $\CS$ theories, 
or line operators in $N=4$ SYM theory. This will be the first type of modification turning out to be crucial 
for our story.

\subsection{Surface operators in class $\CS$ theories} 

By now there exists a fair amount of evidence that
an effective description of a system of M5-M5'-branes obtained by compactification on $C$
is provided by theories of class $\CS$ modified by the presence of a certain type of 
surface operator.  The relevant surface operators can be described by prescribing a certain type of 
singular behaviour of the gauge fields along a two-dimensional submanifold $\CM^2\subset\CM^4$
characterised by a set of parameters $x=(x_1,\dots,x_{3g-3+n})$ if $C$ has genus $g$ and $n$
punctures \cite{AT,FGT}. A generalisation $\CZ^{\rm\sst inst}_\si(a;x,\tau;\ep_1,\ep_2)$ 
of the instanton partition functions can be defined
in the presence of such surface operators, carrying an additional dependence on
the parameters $x$ on top of the dependencies on the scalar zero mode values $a$, the UV
gauge couplings associated to complex structure moduli $\tau$ of $C$, and the 
parameters $\ep_1,\ep_2$ associated to the Omega-deformation. The subscript $\si$ refers to the 
pants decomposition determining the Lagrangian representation being used.
Explicit calculations in \cite{Br,AT,KPPW,Ne,Na} gave evidence for a
generalisation of the AGT-correspondence, identifying $\CZ^{\rm\sst inst}_{\si}(a;x,\tau;\ep_1,\ep_2)$
with the chiral partition functions $\CZ^{\rm\sst WZW}_{\si}(a;x,\tau;k)$ associated to 
conformal blocks of the affine Lie algebras $\hat{\mathfrak{g}}_k$ at level $k=-h^{\vee}-\ep_2/\ep_1$ defined 
by the gluing construction.
The parameters $x$ get identified with coordinates on the moduli space $\mathrm{Bun}_G$ 
of holomorphic $G$-bundles on $C$ under this correspondence.
A proof of this generalisation of the AGT-correspondence
will follow from the result 
announced in \cite{N15}
that $\CZ^{\rm\sst inst}_\si$ satisfies the KZB equations.\footnote{ 
see also {\cite{Ne}} for an earlier result in this direction.}

Our goal in this section will be to generalise the approach of \cite{NW,GW} to theories of class
$\CS$ in the presence of surface operators of co-dimension two. We will propose an effective
description in terms of the two-dimensional sigma model with target $\CM_H(G)$ in which 
the brane  of opers $\mathsf{B}_{\rm op}$ gets replaced by a family of  Lagrangian branes 
$\mathsf{L}_{x}^{\sst (2)}$ supported on the fibers of Hitchin's second fibration over 
bundles $\mathcal{E}_x$ labelled by coordinate $x$ for  $\mathrm{Bun}_G$.
The relevance of the branes $\mathsf{L}_{x}^{\sst (2)}$ in this context and the relation 
to conformal blocks 
has first been proposed by E. Frenkel in \cite{Fr10}, as has been pointed out 
to one of us (J.T.) in 2012. Our goal in the following will be to offer additional support for 
this proposal. 

\subsection{Nahm pole boundary conditions} \label{S:Nahm}

Gaiotto and Witten have classified  1/2 BPS boundary conditions of $N=4$ SYM 
in \cite{GW1}
using the data $(\rho,H,T)$, where $\rho:\mathfrak{sl}_2\ra \mathfrak{g}$ is 
an embedding of $\mathfrak{sl}_2$ into the Lie algebra $\mathfrak{g}$ of the gauge group $G_c$, 
$H$ is a subgroup of the commutant in $G_c$ of the image of $\rho$, and $T$ is a three-dimensional 
SCFT with $N=4$ supersymmetry and at least $H$ global symmetry. We will only need two of the 
simplest of these boundary conditions. 
In the following we will first briefly review the 
so-called Nahm pole boundary condition studied in \cite{GW} which is 
associated to a triple $(\rho,\mathrm{Id},T_\emptyset)$, where $\rho$ is a {\it principal} 
$\mathfrak{sl}_2$-embedding, and $T_\emptyset$ stands for the trivial  three-dimensional 
SCFT. We will then discuss the even simpler case where $\rho$ is replaced by the trivial 
embedding mapping $\mathfrak{sl}_2$ to $0\in\mathfrak{g}$, which will be of particular 
interest for us.

It is for our purposes sufficient to describe 
the Nahm pole boundary conditions for the solutions of the
BPS-equations \cite{KW} characterising field configuration
in $N=4$ SYM preserving certain supersymmetries.  
Restricting attention to solutions to the BPS-equations on $\mathbb{R}\times \mathbb{R}_+ \times C$
which are invariant under translations along $\mathbb{R}$, one gets a system of differential
equations of the form 
\begin{subequations}\label{BPS}
\begin{align}
&[\,\CD_z\,,\,\CD_{\bar{z}}\,]=0\,,\quad [\,\CD_y\,,\,\CD_{{z}}\,]=0\,,
\quad [\,\CD_y\,,\,\CD_{\bar{z}}\,]=0\,,\label{Fterm}\\
&\sum_{i=1}^3[\,\CD_i\,,\,\CD_i^\dagger\,]=0\,. \label{Dterm}\,,
\end{align}
\end{subequations}
where the notations $z=x_2+ix_3$ and $y=x_1$ have been used, and the
differential operators $\CD_i$ are of the form\footnote{Our conventions differ slightly from \cite{GW}.} 
\begin{equation}
\begin{aligned}
&\CD_z=\zeta\pa_z+\mathcal{A}_z\,,\\
&\CD_{\bar{z}}=\pa_{\bar{z}}+\mathcal{A}_{\bar{z}}\,,
\end{aligned}
\qquad
\begin{aligned}
&\mathcal{A}_z=\zeta A_z+\phi_z\,,\\
&\mathcal{A}_{\bar{z}}=A_{\bar{z}}+\zeta\phi_{\bar{z}}\,,
\end{aligned}
\qquad \CD_y=\pa_{y}+A_y-i\phi_y\,.
\end{equation}
The parameter $\zeta$ determines the supersymmetries that are preserved. It is proposed in \cite{GW} that 
the space of solutions to \rf{BPS} modulo compact gauge transformations is isomorphic to the moduli 
space of the solutions to the ``F-term'' equations \rf{Fterm} modulo complex gauge transformations.
Equations  $[\CD_z,\CD_{\bar{z}}]=0$  determine a flat 
complex connection on $C$ at each fixed $y$. 
The remaining equations in \rf{Fterm} imply that the $y$-dependence
of this flat connection is represented by complex gauge transformations. 

Boundary conditions of Nahm pole type are defined in \cite{GW}
by demanding that the solutions to \rf{BPS} have a
singular behaviour of the form 
\begin{equation}\label{Nahm}
\mathcal{A}_z\,\underset{y\ra 0}{\sim}\,\mathfrak{t}_- \,y^{-1}+\CO(y^0)\,,\qquad
\mathcal{A}_{\bar{z}}\,\underset{y\ra 0}{\sim}\,\CO(y^0)\,,\qquad
\mathcal{A}_1\,\underset{y\ra 0}{\sim}\,\mathfrak{t}_3 \,y^{-1}+\CO(y^0)\,,
\end{equation}
with $\mathfrak{t}_+=\mathfrak{t}_1+i\mathfrak{t}_2$, and $\mathfrak{t}_i$, $i=1,2,3$, being the 
generators of a principal $\mathfrak{sl}_2$ subalgebra of $\mathfrak{g}$. 
By a gauge transformation we may always set $\mathcal{A}_{\bar{z}}$ to zero, allowing us to represent
the flat connection on $C$ we get at each $y$ as a local system $(\mathcal{E}_{y},\nabla_{y}')$ consisting of a 
holomorphic bundle and a holomorphic connection $\nabla_{y}'=dz(\pa_z+\mathcal{A}_z(z;y))$.
In the case $\mathfrak{g}=\mathfrak{sl}_2$,  we may reformulate 
the first condition in \rf{Nahm} as the condition that
there exists a basis of sections $s=\{s_1,s_2\}$ with respect to which $\mathcal{A}$ 
has the form $\mathcal{A}=g\tilde{\mathcal{A}}g^{-1}+gd g^{-1}$, with 
\begin{equation}\label{Nahm'}
\tilde{\mathcal{A}}_z\,\underset{y\ra 0}{\sim}\,\bigg(\begin{matrix} 0 & t\\ 1 & 0\end{matrix}\bigg)+\CO(y^1) \,,
\qquad g\underset{y\ra 0}{\sim} \bigg(\begin{matrix} y^{1/2}& 0 \\  0& y^{-1/2} \end{matrix}\bigg) +\CO(y^0).
\end{equation}
Horizontal sections $(d+\mathcal{A})s=0$ will then have a first component $s_1$ vanishing as $y^{1/2}$. 
As explained in \cite{GW}, this implies that the local system 
$\lim_{y\ra 0}^{}(\tilde{\mathcal{E}}_{y},\tilde{\nabla}_{y}')$ on $C$ is an oper. 

The Nahm pole boundary condition has the feature that it breaks $G_c$ 
maximally since the commutant 
of the principal $\mathfrak{sl}_2$-embedding is trivial.
At the opposite extreme, associated to the trivial
$\mathfrak{sl}_2$-embedding, one gets a similar boundary condition 
associated to a triple $(0,\mathrm{Id},T_\emptyset)$ by fixing the boundary 
value of the gauge field $\mathcal{A}_{\bar{z}}$. 

In the reduction to two dimensions having fixed $\mathcal{A}_{\bar{z}}$ at the boundary of $I$ 
implies having fixed a holomorphic bundle on $C$, leaving the $(1,0)$ part of the 
complex gauge field unconstrained. The moduli space $\CM_{dR}(G)$ 
of pairs $(\CE,\nabla_\zeta')$ is isomorphic to the Hitchin moduli space $\CM_H(G)$ via the
non-abelian Hodge (NAH) correspondence$^{(g)}$. Fixing $\CE$ therefore defines a submanifold in $\CM_H(G)$
which is Lagrangian with respect to the holomorphic symplectic form $\Omega_\zeta$,
and holomorphic w.r.t. to the complex structure $I_\zeta$. For $\zeta=i$ one 
has $\Omega_\zeta=\Omega_J$, $I_\zeta=J$, leading to the identification of the 
brane coming from the reduction of the zero Nahm pole boundary condition
as an (A,B,A)-brane in the A-model with the symplectic structure $\omega_I$ 
used in \cite{NW}. 

We are proposing that the zero Nahm pole boundary condition represents the
presence of a surface operator of co-dimension two. Indeed, as was argued in 
\cite{FGT}, the presence of a co-dimension two surface operator naturally 
introduces additional background data which can be geometrically 
represented as the choice of a holomorphic bundle on $C$.

\subsection{Relation to conformal blocks}

The Lagrangian submanifolds $\mathsf{L}_{\CE,\zeta}^{\sst (2)}$ 
of $\CM_H(G)$ defined by considering
pairs $(\CE,\nabla')$ with fixed $\CE$ are sometimes called 
the fibers of Hitchin's second fibration. Picking a
system $x$ of coordinates for $\mathrm{Bun}_G$ we 
will use the notation $\mathsf{L}_{x}^{\sst (2)}$ for $\mathsf{L}_{\CE_x,i}^{\sst (2)}$ 
with $\CE_x$ being a bundle representing the point in $\mathrm{Bun}_G$ 
specified by the coordinates $x$.

{The} results on the generalisation of the AGT-correspondence in the presence of the
surface operators of co-dimension two mentioned above suggest that the space 
$\CH^{\sst (2)}_x={\rm Hom}_{\CM_H(G)}(\mathcal{B}_{\rm cc},\mathsf{L}_{x}^{\sst (2)})$ can be identified with 
the space  of conformal blocks of the affine Lie algebra
$\hat{\mathfrak{g}}_k$ at level $k=-h^{\vee}-\frac{\ep_2}{\ep_1}$ on $C$.  

We need to note, however, that finding the  proper definition of both $\CH^{\sst (2)}_x$ and 
the relevant spaces of conformal blocks is nontrivial in the infinite-dimensional situation
at hand. The usual algebraic definition of conformal blocks defines a space
$\mathcal{CB}_{\rm al}(C,\hfg_k,\mathcal{E}_x)$ that is too large to be
relevant for us.  We'll need to consider a subspace denoted
$\mathcal{CB}_{\rm te}(C,\hfg_k,\mathcal{E}_x)$ of ``tempered'' conformal blocks.
The relation between these two spaces is in some respects similar to the relation 
between spaces $\mathcal{K}_{\rm al}$ of {\it formal} power series $\sum_{n\in\BZ}a_nz^n$ to 
the spaces $\mathcal{K}_{\rm te}$ of tempered 
distributions on the unit circle. The latter can be represented by Fourier series
$\sum_{n\in\BZ}a_ne^{in\si}$ in the distributional sense, leading to an embedding of 
$\mathcal{K}_{\rm te}$ into $\mathcal{K}_{\rm al}$.
However, being tempered imposes growth conditions on the coefficients $a_n$,
making  $\mathcal{K}_{\rm te}$ strictly smaller than $\mathcal{K}_{\rm al}$.

An algebraic counterpart for $\CH^{\sst (2)}_x$, here denoted $\CH^{\sst (2)}_{{\rm al},x}$,
was proposed in the work \cite{Fr10} of E. Frenkel,
where it was proven that $\CH^{\sst (2)}_{{\rm al},x}\simeq \mathcal{CB}_{\rm al}(C,\hfg_k,\mathcal{E}_x)$.
We will argue that a different definition
for $\CH^{\sst (2)}_{x}$ is appropriate in this context, leading to an isomorphism
$\CH^{\sst (2)}_{x}\simeq \mathcal{CB}_{\rm te}(C,\hfg_k,\mathcal{E}_x)$
with a subspace $\mathcal{CB}_{\rm te}(C,\hfg_k,\mathcal{E}_x)$ 
of ``tempered" conformal blocks within $\mathcal{CB}_{\rm al}(C,\hfg_k,\mathcal{E}_x)$.
We can not discuss the 
definition of $\mathcal{CB}_{\rm te}(C,\hfg_k,\mathcal{E}_x)$ fully in this short note,
we plan to return to this point elsewhere. Instead we will
in the following discuss  evidence for $\CH^{\sst (2)}_{x}\simeq \mathcal{CB}_{\rm te}(C,\hfg_k,\mathcal{E}_x)$
coming from the $2d$ sigma model.

\subsubsection{Module structures}

To begin with, let us note that $\CH^{\sst (2)}_x$ has {\it two} natural module structures coming from the 
vertex operators associated to the elements of the algebra
$\mathcal{A}_{\rm cc}^\hbar={\rm Hom}_{\CM_G(C)}(\mathsf{B}_{\rm cc},\mathsf{B}_{cc})$.
For the case of interest one should represent $\CM_H(G)$ as moduli space of flat complex
connections on $C$ using the NAH correspondence, allowing us to consider 
{\it two} algebraic structures coming from the representation of $\CM_H(G)$ as character 
variety$^{(g)}$ $\CM_{B}(G)$ and as moduli space $\CM_{dR}(G)$ of pairs $(\CE,\nabla')$,
respectively. Different algebraic structures determine different sets of basic field 
variables to be used in the definition of the vertex operators in $\mathcal{A}_{\rm cc}$.
One may accordingly distinguish
$\mathcal{A}_{B}^\hbar={\rm Hom}_{\CM_B(C)}(\mathsf{B}_{\rm cc},\mathsf{B}_{cc})$
and
$\mathcal{A}_{dR}^\hbar={\rm Hom}_{\CM_{dR}(C)}(\mathsf{B}_{\rm cc},\mathsf{B}_{cc})$,
where $\hbar=\ep_1/\ep_2$.

On the one hand it was argued in \cite{NW} that the algebra $\mathcal{A}_{B}^\hbar$ can be identified with the 
quantised algebra of functions on  $\CM_{B}(G)$. 
The arguments in  \cite{Ka08} may, on the other hand, be applied to the situation 
studied in \cite{NW} leading to the conclusion that  
$\mathcal{A}_{dR}^\hbar\simeq \CD_{\hbar}$, 
the quantised algebra of functions on  $\CM_{dR}(G)$, 
which may be identified with the algebra 
of differential operators on a certain line bundle $\CL^{-h^\vee-\ep_2/\ep_1}$ over $\mathrm{Bun}_G$.

It well-known that the space 
$\mathcal{CB}_{\rm al}(C,\hat{\mathfrak{g}}_k,\mathcal{E}_x)$ of conformal blocks of the affine Lie algebra
$\hat{\mathfrak{g}}_k$ at level $k=-h^{\vee}-\frac{\ep_2}{\ep_1}$ has a module 
structure with respect to $\mathcal{A}_{dR}^{\hbar}$. 
The $\mathcal{A}_{dR}^{\hbar}$-module structure is a direct consequence 
of the defining Ward identities. 

A $\mathcal{A}_{B}^{\hbar}$-module structure can defined on spaces of conformal blocks
by using degenerate
representations of $\hat{\mathfrak{g}}_k$ to define analogs of the Verlinde line operators \cite{AGGTV,DGOT}
in this case.  It will be shown elsewhere that the algebra
of Verlinde line operators on affine Lie algebra conformal blocks 
may be identified with  $\mathcal{A}_{B}^{\hbar}$.
The known definitions work with conformal blocks constructed using the gluing 
construction. Such conformal blocks have particularly nice properties, one of which being that 
the canonical connection may be
integrated over all of $\mathrm{Bun}_G^{\rm vs}$, the subset of $\mathrm{Bun}_G$ 
containing the ``very stable'' bundles not admitting a nilpotent Higgs field which
is not the case for generic 
elements of $\mathcal{CB}^{\rm al}(C,\hat{\mathfrak{g}}_k,\mathcal{E}_x)$. This suggests 
that the $\mathcal{A}_{B}^{\hbar}$-module structure can only be defined on suitable  subspaces 
$\mathcal{CB}_{\rm te}(C,\hat{\mathfrak{g}}_k,\mathcal{E}_x)$ of ``tempered'' conformal blocks.

The existence of two module structures on $\CH^{\rm (2)}_x$ can be regarded as a first piece of evidence for the 
conjectured isomorphism
$\CH^{\rm (2)}_x\simeq\mathcal{CB}_{\rm te}(C,\hat{\mathfrak{g}}_k,\mathcal{E}_x)$. 
Further evidence will be given below.

\subsubsection{Relation to $\bar\partial$-cohomology}

To begin with, let us return to the case studied by Nekrasov and Witten in \cite{NW}.
We first note that the space ${\rm Hom}_{\CM_H(G)}(\mathsf{B}_{\rm cc}, \mathsf{B}_{\rm op})$ has a realisation in the 
A-model with symplectic structure $\omega=\omega_I$. The boundary conditions on the strip are 
given as the canonical coisotropic A-branes  $\mathsf{B}_{\rm cc}$ and the Lagrangian A-brane
$\mathsf{B}_{\rm op}$. The Chan-Paton curvature $F$ is given by the 
symplectic form $\omega_J$ on Hitchin moduli space. It follows that the
complex structure determined by the canonical coisotropic A-brane  $\mathsf{B}_{\rm cc}$ is 
$\omega^{-1}F=K$. It is argued in \cite{NW} that in this case the space
${\rm Hom}_{\CM_H(G)}(\mathsf{B}_{\rm cc}, \mathsf{B}_{\rm op})$ can be identified with the space of 
holomorphic sections of the line bundle $K^{{1}/{2}}_{\rm op}$ on the 
subspace $\mathrm{Op}_{\fsl_2}(C)$ of opers in $\CM_{dR}(G)$. Indeed, the space 
${\rm Hom}_{\CM_H(G)}(\mathsf{B}_{\rm cc}, \mathsf{B}_{\rm op})$ will have a realisation as space of 
functions of the zero modes of the A-model on the strip. We have Neumann-type boundary conditions
for the sigma model fields representing coordinates on $\mathrm{Op}_{\fsl_2}(C)$
on both ends of the strip, while the remaining 
fields have Dirichlet-type boundary conditions on one end only. The zero modes of the A-model may therefore
be represented in terms of coordinates on $\mathrm{Op}_{\fsl_2}(C)$. Standard arguments briefly reviewed in \cite{NW} may then 
be used to identify the space of supersymmetric ground states of the sigma model on the strip with the 
$\bar{\partial}$-cohomology with values in $K^{{1}/{2}}_{\rm op}$. 

%An equivalent description is obtained by mirror symmetry, allowing us to represent 
%$\CH$ as ${\rm Hom}(\mathsf{B}_{\sst N},\mathsf{B}_{\rm cc})$. We may still
%use a representation as a A-model with symplectic structure $\omega=\omega_I$, but 
%the canonical coisotropic brane now has Chan-Paton curvature $F=\omega_K$. 
%Following the same reasoning as before leads to the identification of 
%$\CH$ with the space of sections of the line bundle 
%$K^{{1}/{2}}_{\sst N}$ which are holomorphic in complex structure $J=\omega^{-1}F$
%rather than $K$. 

\subsubsection{Relation to conformal blocks: Virasoro case}\label{HomVirConf}

Note that $\mathrm{Op}_{\fsl_2}(C)$ is (non-canonically) isomorphic to the complex vector space $H^0(C,K^2_C)$
which is topologically trivial. The space $H^0(C,K^2_C)$ is canonically isomorphic to the 
cotangent fiber $T^*\mathcal{T}(C)|_{C}$ to the Teichm\"uller space $\mathcal{T}(C)$.
It will be important for us to observe that there is a map from the cotangent space 
$T^*\mathcal{T}(C)|_{C}$
to the Teichm\"uller space $\mathcal{T}(C)$ relating the natural complex structures on these
spaces.

In order to introduce the complex structure on $\mathcal{T}(C)$ one may describe the 
Teichm\"uller variations in terms of the %so-called 
harmonic Beltrami differentials, which are
$(-1,1)$-forms $\mu$ satisfying $\pa_z(\eta\mu)=0$, where $\eta$ is the hyperbolic metric 
uniquely determined by the complex structure on $C$. The
complex structure on the vector 
space of harmonic Beltrami differentials thereby defines the complex structure on $\mathcal{T}(C)$. 
To a holomorphic $(2,0)$-form $\theta$ one may assign the corresponding 
harmonic Beltrami differential $\eta^{-1}\bar{\theta}$. This defines a
complex anti-linear map from $H^0(C,K^2_C)$ to 
the space of harmonic Beltrami differentials expressing the 
relation between the natural complex structures on $\mathcal{T}(C)$ and on $H^0(C,K^2_C)$, respectively.
%\footnote{
%The map from Beltrami differentials $\mu$ to quadratic differentials $\eta\bar{\mu}$
%coincides with the map from holomorphic tangent to 
%anti-holomorphic cotangent spaces defined by the Weil-Petersson metric. 
%The complex structure on  $\mathcal{T}(C)$ defined in this way coincides with the one provided by 
%Kodaira-Spencer theory thanks to Dolbeault's theorem. This is one way to  
%see in which respect the  Weil-Petersson metric is distinguished.}
It allows us to identify the space of holomorphic functions on $\mathrm{Op}_{\fsl_2}(C)$ with the space 
${\rm Fun}_{\rm hol}(\mathcal{T}(C))$ 
of (anti-) holomorphic functions on $\mathcal{T}(C)$. 

The spaces ${\rm Fun}_{\rm hol}(\mathcal{T}(C))$, on the other hand,
are known to be isomorphic 
with (sub-)spaces of Virasoro conformal blocks. 
Conformal blocks on closed\footnote{We temporarily restrict to this case to simplify the exposition.} surfaces $C$ are defined as linear functionals $f$ on  the vacuum representation $V_0$
of the Virasoro algebra
satisfying the conformal Ward identities describing invariance of $f$ under the natural action of the 
algebra of vector fields holomorphic away from a  point. The vector space of conformal blocks carries 
a canonical projectively flat connection defined by means to the Virasoro action on $V_0$. 
The curvature of the canonical connection may be trivialised locally on the moduli space $\CM_{g}$ of 
complex structures. On suitable subspaces of the space of all algebraically defined 
conformal blocks one may integrate the resulting
flat connection to define horizontal sections $f_{\tau}$ on open subsets of $\CM_{g}$ with 
local coordinates $\tau$. The values of 
the conformal blocks  $\CZ_f(\tau)=f_\tau(v_0)$ on the  highest weight vector $v_0$ 
are called  chiral partition functions. The 
Virasoro uniformization theorem and conformal Ward identities
relate the derivatives of $\CZ_f(\tau)$  to the data characterising
the conformal blocks $f$. One thereby gets a one-to-one correspondence between (locally defined) 
functions $\CZ(\tau)$ and ``integrable'' conformal blocks $f$. This correspondence is essentially canonical:
The Virasoro uniformisation describes the local structure of 
$\CM_{g}$, and this is encoded in the definition of the conformal blocks.

%The arguments used to reach the same conclusion in  \cite{NW} were based on the 
%comparison of the natural action of the quantised algebra of functions on $\CM_{B}(G)$ on 
%${\rm Hom}(\mathsf{B}_{\sst N},\mathcal{B}_{\rm cc})$ with the 
%action of the Verlinde loop operators on spaces of Virasoro conformal blocks. 
%The line of arguments above offers 
%a more direct way to arrive at this conclusion.

\subsubsection{Relation to conformal blocks: Kac-Moody case}

It is possible to argue that $\CH_x^{\sst (2)}$
is isomorphic to the space $\mathcal{CB}(C,\hat{\mathfrak{g}}_k,\CE_x)$ of 
conformal blocks for $\hat{\mathfrak{g}}_k$ twisted by the bundle $\CE_x$ in a similar way as 
was argued in Section \ref{HomVirConf}.
Following the the same reasoning as described there one would arrive at the 
conclusion that $\CH_{\mathcal{E}}$ is isomorphic to the space of holomorphic functions
on $H^0(C,{\rm End}(\mathcal{E})\otimes K_C)$. The space 
$H^0(C,{\rm End}(\mathcal{E})\otimes K_C)$ is the holomorphic 
cotangent space $T^*{\rm Bun}_G$. 
There exists a canonical hermitian metric on $T{\rm Bun}_G$ which is analogous to the 
Weil-Petersson metric in Teichm\"uller theory, defined using the
Narasimhan-Seshadri theorem as analog of the uniformisation theorem.
This metric identifies
the holomorphic co-tangent space with the anti-holomorphic tangent space, 
relating the natural complex structures
on $H^0(C,{\rm End}(\mathcal{E})\otimes K_C)$ and 
$T{\rm Bun}_G$.

The space of holomorphic functions on $T{\rm Bun}_G$ embeds naturally into 
$\on{Jet}_\mathcal{E}$, the space of functions on a 
formal neighbourhood of $\mathcal{E}$ in ${\rm Bun}_G$. 
It remains to notice
that $\on{Jet}_\mathcal{E}$ is isomorphic to the 
space of conformal blocks, 
$\on{Jet}_\mathcal{E}\simeq\mathcal{CB}_{\rm al}(C,\hat{\mathfrak{g}}_k,\CE_x)$ \cite[Section 18.2]{FBZ}. 
This observation was used by E. Frenkel in \cite{Fr10} to prove the
algebraic counterpart 
$\CH^{\sst (2)}_{{\rm al},x} \simeq \mathcal{CB}_{\rm al}(C,\hfg_k,\CE_x)$ of our claim.
It follows that spaces of
holomorphic functions on $T{\rm Bun}_G$ can be identified with subspaces of 
"well-behaved" conformal blocks within $\mathcal{CB}(C,\hat{\mathfrak{g}}_k,\CE_x)$,
further supporting the conjecture 
$\CH^{\sst (2)}_{x} \simeq \mathcal{CB}_{\rm te}(C,\hfg_k,\CE_x)$
proposed above.

%{\tt To do:}  Singular locus of corresponding
%$\CD$-module? Automatically restricted to  open 
%subset of ``very stable'' bundles?

\section{Partition functions versus conformal blocks}\label{partfct}

Partition functions play a central role in the AGT-correspondence and its generalisations,
while they do not appear in the approach of Kapustin and Witten 
to the geometric Langlands correspondence. In order to develop a unifying 
framework we will now discuss how to define the relevant 
partition functions in the effective description furnished by the 
two-dimensional sigma models.

\subsection{TQFT set-up in four and two dimensions}

Using the reduction of class $\CS$-theories to the two-dimensions we will
in the following motivate  a description of the  partition functions within the two-dimensional sigma model with target 
$\CM_H(G)$. It will be based on yet another type of boundary condition 
denoted $\mathsf{L}^{\sst (1)}_{a}$. The notation $\mathsf{L}^{\sst (1)}_{a}$ 
is motivated by a link to Hitchin's {\it first} fibration 
which will be disccussed later.

%\subsubsection{Four-dimensional set-up}

Following \cite{NW} we will consider topologically twisted class $\CS$-theories on hemispheres 
$B_{\ep_1\ep_2}^4$ with Omega-deformation.
The topologically twisted class $\CS$-theory associates a vector space 
$\CH_{\rm top}=Z(M^3_{\ep_1\ep_2})$ to $M^3_{\ep_1\ep_2}=\pa B^4_{\ep_1\ep_2}$, here identified with the 
cohomology of $Q$, the supercharge 
that can be preserved on $B_{\ep_1\ep_2}^4$.
One may use the 
path integral over the 4d hemisphere $B_{\ep_1\ep_2}^4$ to define a vector $\Psi 
\in\CH_{\rm top}$. Wave-functions $\Psi(a)$
of the vector $\Psi$ may be identified with the 
partition functions $Z(B^4_{\ep_1,\ep_2};\mathcal{B}_a)$ 
defined by imposing suitable $Q$-invariant boundary conditions $\mathcal{B}_a$ labelled by parameters $a$
at $M^3_{\ep_1\ep_2}$. 
Such boundary conditions can be identified with the boundary conditions 
at the infinity of $\BR^4_{\ep_1\ep_2}$ used to define the Nekrasov partition functions, fixing in particular
the zero modes of the scalars in the vector multiplets to have values collected
in the vector $a=(a_1,\dots,a_{3g-3+n})$.
The boundary conditions $\mathcal{B}_a$ 
define a family of boundary states $\beta_a$, allowing us to represent
$Z(B^4_{\ep_1,\ep_2};\mathcal{B}_a)$ as an overlap $\langle\beta_a,\Psi\rangle$.

%\subsubsection{2d reductions}

In the reduction of the class $\CS$-theory to a $2d$ topological sigma model 
one should get the following representation of the 4d TQFT data introduced above:
\begin{itemize}
\item The vector space  $\CH_{\rm top}\simeq Z(S^3_{\ep_1,\ep_2})\ra Z(I)$.
\item The vector $\Psi=Z(B^4_{\ep_1,\ep_2})\ra Z(T_{\ep_1,\ep_2})\in\CH_{\rm top}$, where 
$T_{\ep_1,\ep_2}$ is the open triangle with ``upper'' side  removed, 
topologically equivalent to $\BR_-\times I$, partially 
compactified by adding a point at the infinity of $\BR_-$. The boundary of $T_{\ep_1,\ep_2}$ is $\{0\}\times I$.
\item The partition function $Z(B^4_{\ep_1,\ep_2};\mathcal{B}_a)\ra Z(T_{\ep_1,\ep_2};\mathsf{B}_a)$
gets associated to a triangle $T_{\ep_1,\ep_2}$ with a
boundary condition $\mathsf{L}_{a}^{\sst (1)}$ assigned to the 
upper side $\{0\}\times I$. $\mathsf{L}_{a}^{\sst (1)}$ is 
defined  from the boundary condition $\mathcal{B}_a$
assigned to $M^3_{\ep_1\ep_2}$ by the reduction to one 
dimension.
\end{itemize}
This means that the instanton partition functions $\CZ(a;x;\tau;\ep_1,\ep_2)$
get represented by partition functions of the sigma model
on a triangle which has sides coloured by $(\mathsf{B}_{\rm cc}, \mathsf{L}_a^{\sst (1)}, \mathsf{L}_x^{\sst (2)})$.

\subsection{$\CD$-modules versus partition functions}

States $\psi$ in quantum theory are  abstractly 
represented by vectors in a 
Hilbert space $\CH$. A concrete representation as space of 
wave-functions $\psi(x)$ is obtained by introducing a family of 
elements $\de_x$ of the (hermitian) dual $\CH^\dagger$ of $\CH$ allowing us to 
represented $\psi(x)$
as $\langle\de_x,\psi\rangle$, with $\langle . , .\rangle:\CH^{\dagger}\times\CH\ra \BC$ 
being the  natural pairing.
In the present context we may interpret the partition functions
$\CZ=\CZ(a;x,\tau;\ep_1,\ep_2)$ as wave-functions $\phi_x(a)=\langle\de_a^{\sst(1)},\phi_x^{}\rangle$
of an element $\phi_x\in \CH_{x}^{\sst(2)}:=
{\rm Hom}_{\CM_H(G)}^{}(\mathsf{B}_{\rm cc}^{},\mathsf{L}_{x}^{\sst(2)})$ created
by the sigma model path integral over the semi-infinite strip $\BR_-\times I$.
The ``boundary state'' $\de_a^{\sst(1)}$ represents the boundary 
condition $\mathsf{L}_{a}^{\sst(1)}$  %imposed 
at $\{0\}\times I$.

The overlap $\langle\de_a^{\sst(2)},\phi_x^{}\rangle$ has a representation as a 
partition function in the topological sigma model on a triangle 
with three different types of boundary conditions
$(\mathsf{B}_{\rm cc},\mathsf{L}_{x}^{\sst(2)},\mathsf{L}_{a}^{\sst(1)})$
 assigned to the 
three sides. Time-reversal symmetry identifies the 
complex conjugate $\CZ^*$ of  $\CZ$ with a 
partition function in the topological sigma model on a triangle 
with boundary conditions 
$(\mathsf{B}_{\rm cc}^{},\mathsf{L}_{a}^{\sst(1)},\mathsf{L}_{x}^{\sst(2)})$
appearing in a different order. 
$\CZ^*$ admits a Hamiltonian interpretation as
a wave-function $\vf_a(x)=\langle\de_x^{\sst(2)},\vf_a^{}\rangle$
of an element $\vf_a\in \CH_{x}^{\sst(1)}:=
{\rm Hom}_{\CM_H(G)}^{}(\mathsf{B}_{\rm cc}^{},\mathsf{L}_{a}^{\sst(1)})$ created
by the sigma model path integral over the semi-infinite strip $\BR_-\times I$.

Both spaces $\CH_{a}^{\sst(1)}$ and $\CH_{x}^{\sst(2)}$ have a $\mathcal{A}_{dR}^{\hbar}$-module 
structure coming from the natural action of the algebra $\mathcal{A}_{dR}^{\hbar}$ 
on ${\rm Hom}_{\CM_H(G)}^{}(\mathsf{B}_{\rm cc}^{},\mathsf{B})$. We have argued above 
that this module  structure coincides with the one on spaces of conformal blocks
following from the affine algebra Ward identities. The resulting
structure as twisted $\CD$-module coincides with the one
defined by the KZB equations
satisfied by the chiral partition functions 
corresponding to the conformal blocks of affine Lie algebras.
The partition functions
$\CZ$ are solutions to the KZB equations. The $\CD$-modules 
$\CH_{a}^{\sst(1)}$ and 
$\CH_{x}^{\sst(2)}$ can be fully characterised by the corresponding spaces of 
wave-functions, physically represented as the partition functions $\CZ$.

In the limit $\ep_2\ra 0$ discussed below we will propose that
$\CH_{{a}}^{\sst (1)}$  appears in the Kapustin-Witten approach to the 
geometric Langlands correspondence while $\CH_x^{\sst(2)}$ appears to be a more natural
space to consider  in the context of generalisations of the AGT correspondence.
The relation between the two Hamiltonian interpretations of the partition function $\CZ(x,{a},\tau;\ep_1,\ep_2)$ 
following from an open TQFT version of modular invariance
represents a crucial link between these two stories.

\subsection{Sigma model description of the brane $\mathsf{L}_{a}^{\sst(1)}$} \label{branes1}

Let us now propose a description of the boundary condition $\mathsf{L}_{a}^{\sst(1)}$ as a
Lagrangian submanifold in $\CM_H(G)$. For future use we will include the label $\si$
referring to the Lagrangian representation of class $\CS$ theories 
used to define the branes $\mathsf{L}_{a}^{\sst(1)}$ into the notation, changing it into 
$\mathsf{L}_{\si,a}^{\sst(1)}$. We propose that 
the branes $\mathsf{L}_{\si,a}^{\sst(1)}$ can be described as the 
orbits in the character variety $\CM_B(G)$ 
of the Hamiltonian flows generated by the 
collection of  length functions $a=(a_1,\dots,a_{3g-3+n})$ representing a 
Poisson-commuting half of 
the system of complex Fenchel-Nielsen coordinates{$^{g)}$} used in a related context in \cite{NRS}.
The orbits are mapped via holonomy map and 
the  NAH correspondence to the Hitchin moduli space $\CM_H(G)$.
The branes $\mathsf{L}_{{\ep},a}^{\sst(1)\si}$
are natural objects from the point of view of the integrable structure of the character variety
$\CM_B(G)$.
Different systems of complex 
Fenchel-Nielsen coordinates are labelled by pants decompositions $\si$.

Our proposal is based on the known relations {\crd \cite{DMO}} between Wilson loop observables in class $\CS$ 
theories and co-dimension four defects in the six-dimensional $(2,0)$ theory supported on 
products of two circles $S^1_\CS$ and $S^1_C$ in  $\CM^4$  and $C$, respectively. These
relations carry over to the case where surface operators are present. Restricting attention 
to Wilson loops not sharing the support of the surface operators in $\CM^4$ one may argue
as before that the eigenvalues of Wilson loop observables can be expressed in terms 
of the scalar zero modes $a$ appearing in the arguments of the instanton partition functions 
$\CZ({\crd a;x},\tau;\ep_1,\ep_2)$. 

One may alternatively adapt the arguments in 
\cite{AGGTV} to the case with additional surface operators of 
codimension two, which can be based on the results described in 
Section \ref{Hecke1}.

\section{Another type of surface operators} 

One of the key ingredients in the geometric Langlands correspondence is the Hecke eigenvalue 
property. In order to prepare for the discussion of the Hecke eigenvalue property in the 
following section we will now discuss another type of surface operators giving 
us a physical realisation of the Hecke functors.

\subsection{String-theoretical motivation}

From the perspective of the six-dimensional $(2,0)$ theory it seems
natural to investigate the effects of modifying the theory by defects of 
co-dimension four. This defines natural probes of the 
six-dimensional $(2,0)$ theories and their dimensional reductions
which have already been investigated extensively in the 
context of the AGT correspondence \cite{AGGTV}. We will
now discuss briefly how the picture gets modified if 
one considers surface operators of  co-dimensions two and four 
at the same time. The existence of such systems is suggested 
by string theory. Following the discussion in \cite{FGT} one may,
in particular, consider a configuration of 
M5-M5'-M2-branes in M-theory having support indicated in the
following table:
\begin{center}
\begin{tabular}{l || c|c|c|c|c|c|c|c|c|c|c}
Brane & 0 & 1 & 2 & 3 & 4 & 5 & 6 & 7 & 8 & 9 & 10 \\ \hline\hline
M5  & x & x & x & x &   &   & x &   &   &   & x  \\
M5' & x & x &   &   &   &   & x &   & x & x & x \\
M2  & x & x &   &   &   &   &  &  x &   &   &   \\
% M2' & x & x &   &   &   &   &  &   & x &  &  \\
\end{tabular}
\end{center}
Having support of this type means that the co-dimension four surface operators   
on $\CM^4\times C$, with $\CM^4$ a 
fibered product locally of the form\footnote{The lower subscripts in the notation 
$S_i^1$ refer to the rotation symmetry used to define the Omega-deformation with
parameter $\ep_i$.} $\BR\times I\times
S^1_1\times S^1_2$
will be represented as 
\begin{itemize}
\item
surface operators on $\CM^4$ supported on $\CM^2\simeq \BR\times S_{1}^1$,
\item
line operators in $N=4$ SYM on $\BR\times I\times C$, 
supported on $\BR\times \{\pi\}\times \{P\}$,
\item
or line operators in complex {\crd CS} theory on $\BR\times C$, 
supported on $\BR\times \{P\}$.
\end{itemize}
It easily follows from these observations that the presence of such co-dimension four surface
operators modifies the boundary conditions in the two-dimensional sigma model 
describing the reduction to $\BR\times I$. 
We will now briefly describe the modifications that result from the presence of 
surface operators of co-dimension four in the effective description as two-dimensional CFT.

\subsection{Resulting modification of the (generalised) AGT-correspondence}

It has been argued in \cite{AGGTV} that the presence of
co-dimension four surface operators in the six-dimensional $(2,0)$ theory
is described after compactification on $C$ as a system where
a $4d$ theory of class $\CS$ is coupled to certain two-dimensional 
gauged linear sigma models (GLSM) on a two-dimensional surface $\CM^2$ 
in $\CM^4$. The characteristic feature of the 
relevant GLSM is the infrared description in terms of nonlinear 
sigma model with  Grassmannians targets. In the simplest case, 
$\mathfrak{g}= \mathfrak{sl}_2$, one gets, for example, the 
GLSM flowing to the $\mathbb{CP}^1$ sigma model.

Our main conjecture concerning the realisation of co-dimension four surface operators in
the presence of co-dimension two surface operators within CFT is the following:
\begin{equation}\label{Heckeconj}
\boxed{ 
\quad
\begin{aligned}
&\text{Addition of a surface operator of co-dimension four at $P\in C$}\\
& \text{gets represented by an insertion of the 
$\widehat{\mathfrak{sl}}_{2,k}$-module $W_{1/2\hbar}$ at $P$.}
\end{aligned}
\quad
}
\end{equation}
To list the available support for this conjecture, let us start by noticing that it is 
suggested  by a slight generalisation  of the picture described in \cite{FGT} . 
It was observed in this reference that a system composed out of 
M5-M5'-branes displays an interesting IR duality relating it to a system 
of M5-M2-branes, with a collection of M2-branes located at various points 
on $C$. This
IR duality follows from a variant of the Hanany-Witten brane creation
effect describing a family of IR-equivalent effective theories obtained by 
separating the M5-M5'-branes present in the configuration above along the $x_7$-axis. 
In order to get an IR-equivalent effective description one needs to introduce 
additional $M2$-branes suspended between the branes getting separated, 
and extended along the $x_7$-axis.
The end-points of the $M2$-branes  define a divisor $\sum_{k}P_k$ on $C$.
It was argued in \cite{FGT} that this  IR duality is expressed in
relations between supersymmetric partition functions which get  by 
generalisations of the AGT-correspondence translated into 
known relations between the conformal blocks of the affine algebra 
$\widehat{\mathfrak{sl}}_{2,k}$ and $\mathfrak{Vir}_c$. The relevant conformal blocks of $\mathfrak{Vir}_c$ 
are defined using insertions of additional degenerate representations $V_{-1/2b}$ with $b^2=\hbar$ at
the points $P_1,\dots,P_h$, $h=4g-4+n$. The  %resulting relations reproduce known 
relations between these two types of conformal blocks
follow from a  generalisation of Sklyanin's separation of variables (SOV) method. 

A simple variant of the set-up considered in \cite{FGT} is to modify the 
initial system of M5-M5'-branes by having an additional ``primordial'' M2-brane 
localised at a point $P_0\in C$ in the background, and extended along the $x_7$-axis.
In the IR dual description one would then get an additional degenerate representation 
$V_{-1/2b}$ inserted at $P_0$ on top of the insertions at $P_1,\dots,P_h$.
As explained in \cite{T10}, this is nothing but the representation for 
conformal blocks of the affine algebra $\widehat{\mathfrak{sl}}_{2,k}$
obtained by the SOV when there is an additional insertion of 
the $\widehat{\mathfrak{sl}}_{2,k}$-module $W_{1/2\hbar}$ at $P$.
  
\subsection{Further support}

An alternative argument can be obtained by a suitable generalisation of the arguments 
leading \cite{AGGTV} to describe the effect of additional M2-branes 
by the insertion of degenerate fields.  Addition of such 
a vertex operator modifies the space of instanton partition functions by 
tensoring it with the two-dimensional vector space of vacua of the $\mathbb{CP}^1$
sigma model.  Variation of the position of the co-dimension four surface operator
generates a monodromy representation on this two-dimensional vector space. Taking the 
limit $\ep_2\ra 0 $ results in a partial decoupling of the degrees of freedom on the surface operator
and in the bulk, leading to a factorisation of the instanton partition functions into a four-dimensional
singular, and a two-dimensional regular part. The $\widehat{\mathfrak{sl}}_{2,k}$-module $W_{1/2\hbar}$
is the unique candidate for an insertion in affine Lie algebra conformal blocks that has the
properties (i) to generate a two-dimensional monodromy representation, and (ii) insertions of 
$W_{1/2\hbar}$ factor
off the conformal blocks in the critical level limit. In the following section we will describe 
some elements of this line of arguments in more detail.

It would be nice to check our conjecture \rf{Heckeconj} by localisation calculations, 
generalising the results for codimension 2 surface operators cited above on the one hand, 
and the known results for codimension 4 surface operators (see \cite{GLPP} 
and references therein) on the other hand.

In the reduction to $N=4$ SYM one would represent the co-dimension four surface operators
in terms of the Wilson and 't Hooft line operators in $N=4$ SYM \cite{Wi12}. The relation of these
line operators to the Hecke functors of the geometric Langlands correspondence was extensively 
discussed in \cite{KW}. 
In the  reduction scheme studied by Cordova and Jafferis in \cite{CJ} one will find the Wilson lines
of complex Chern-Simons theory, which should give an alternative view on the 
conjecture formulated in \rf{Heckeconj}. 

One may note, finally, that the Hecke functors in the work of Beilinson and Drinfeld
are described by the critical level limit of insertions of the modules  $W_{-(k+h^{\vee})\la}$
of $\mathfrak{g}_k$ into conformal blocks, 
where $\la$ is the weight of a finite-dimensional representation
of ${}^LG$, see \cite[Section 20.5]{FBZ} for a review. The modules appearing 
in \rf{Heckeconj} are associated to the case where $\la=1/2$
corresponding  to the fundamental representation  of 
${}^LG=\mathrm{PSL}(2,\BC)$.

%\subsection{Reduction to two dimensions - TQFT}

%\begin{itemize}
%\item Presence of surface operators modifies brane $\FB'$  $\Rightarrow$ modifies ${\rm Hom}(\FB_{\rm cc},\FB')$
%\item Using approach of \cite{FGT}:
%$2d$ $M_H$-sigma model: $M2\rightsquigarrow$ Modifications of branes: opers with apparent singularities
%\end{itemize}

%\section{Overview}

\section{Recovering the geometric Langlands correspondence}\label{Hecke1}

Our goal is to elaborate on the relations between generalised versions of the AGT-correspondence
in the presence of surface operators 
to the geometric Langlands correspondence. We will argue that a
direct relation emerges in the Nekrasov-Shatshavili limit $\ep_2\ra 0$. 
To this aim we will first observe that the asymptotics of the 
partition functions $\CZ_\si(a;x,\tau;\ep_1,\ep_2)$ contains solutions 
of the Hitchin eigenvalue equations describing the $\CD$-module structures
in the geometric Langlands correspondence. We will then discuss how the 
crucial Hecke eigenvalue property of the relevant $\CD$-modules 
can be understood from the point of view of the Nekrasov-Shatshavili limit
of class $\CS$ theories with surface operators. Possible interpretations of these
phenomena in term of the $2d$ sigma model with target $\CM_H(G)$
will be discussed in the next section.

\subsection{Nekrasov-Shatashvili limit}

The limit $\ep_2\ra 0$ is called Nekrasov-Shatashvili limit following \cite{NS}. 
The relation $k=-h^{\vee}-\ep_2/\ep_1$ identifies it with the critical level limit in CFT.
In CFT is is known \cite{T10,FGT} that the chiral partition functions of affine Lie algebra conformal blocks 
behave in the critical level limit as
\begin{equation}\label{NSlim}
\mathcal{Z}_\si(a;x,\tau;\ep_1,\ep_2)\sim
e^{-\frac{1}{\ep_2}\mathcal{Y}_\si(a;\tau;\ep_1)}
\Psi(a;x,\tau;\ep_1)\,\big(1+\CO(\ep_2)\big)\,,
\end{equation}
where 
\begin{itemize}
\item $\mathcal{Y}_\si(a;\tau;\ep_1)$ 
is the generating function for the Lagrangian submanifold ${\rm Op}_{\fsl_2}(C_\tau)$ of opers, 
considered in a related context in \cite{NRS},
with  $a$ being the complex Fenchel-Nielsen
coordinates for ${\rm Op}_{\fsl_2}(C_\tau)$ defined from the traces of holonomies along a set of 
curves on $C$ defining the
pants decomposition $\si$.
\item $\Psi(a;x,\tau;\ep_1)$ is an eigenfunction of the Hitchin Hamiltonians, $D_r\Psi=E_r\Psi$, with 
eigenvalues $E_r=E_r(a;\tau;\ep_1)$ obtained from $\mathcal{Y}_\si(a;\tau;\ep_1)$ as 
$E_r=\partial_{\tau_r}\mathcal{Y}_\si(a;\tau;\ep_1)$.
\end{itemize}
The behaviour \rf{NSlim} can be seen as a concrete manifestation of 
 the geometric Langlands correspondence: The points in the character variety 
having complex Fenchel-Nielsen coordinates $(a,\pa_a\mathcal{Y}_\si(a;\tau;\ep_1))$ may be
represented as the holonomies of the $\ep_1$-opers $(\CE_{o},\nabla'_{\ep_1,u})=
 \big(\CE_{o},\big(\ep_1\pa_z+ \big(
\begin{smallmatrix} 0 & u \\ 1 & 0\end{smallmatrix}\big)\big)dz\big)$.
To an oper with $u=u^{\sharp}+\sum_{r=1}^{3g-3+n}E_r\vartheta_r(x)$, where $u^\sharp$ defines a fixed 
reference oper,
and $\vartheta_r(x)(dz)^2$, $r=1,\dots,3g-3+n$ is a basis for $H^0(C,K^2)$, 
we may associate the system of equations $D_r\Psi=E_r\Psi$, $r=1,\dots,3g-3+n$, defining
the $\CD$-module associated to the oper $(\CE_{o},\nabla'_{\ep_1,u})$
by the geometric Langlands correspondence.
  
A behaviour of the type \rf{NSlim}  is easily understood from the point of view of class $\CS$-theories 
with surface operators. 
In the limit $\ep_2\ra 0$ one would expect that a modification of the  boundary conditions
for the gauge fields along a submanifold $\CM^2$ that stays effectively compact when $\ep_2\ra 0$
can not affect the leading singular behaviour  in this limit.
%The latter
%should be described by a two-dimensional TQFT in an $\Omega$-background with 
%parameter $\ep_1$. As $\BR^2_{\ep_1}\times \BR^2$ is topologically equivalent 
%to $B^2_{\ep_1}\times \BR^2$, with $\pa B^2_{\ep_1}\simeq S^1_{\ep_1}$,
%we expect to be 

%\section{Hecke eigenvalue property}\label{Hecke2}

\subsection{Hecke eigenvalue property in CFT}\label{Hecke-CFT}

In the presence of both types of surface operators one
may define further generalisations of the partition functions
$\CZ_v(a;x,\tau;t;\ep_1,\ep_2)$
depending in addition to all the variables considered 
previously on a variable $t$ which may be identified with the coordinate
of a point on $C$, the end-point of an ``primordial'' M2-brane.  We have argued above 
that such partition functions get represented in CFT as chiral partition 
functions of conformal blocks with an additional insertion of a degenerate
field at $t$. The extra label $v$ refers to the elements of a basis for the space of 
vacua of the GLSM providing a $4d+2d$ Lagrangian description. Considering a
fixed choice of pants decomposition we will temporarily suppress the label $\si$.

From CFT \cite{T10,FGT} we know that the 
partition functions $\CZ_v(a;x,\tau;t;\ep_1,\ep_2)$ behave in the limit $\ep_2\ra 0$ as
\begin{equation}\label{NSlim+}
\CZ(a;x,\tau;t;\ep_1,\ep_2)\sim
e^{-\frac{1}{\ep_2}\mathcal{Y}(a;\tau;\ep_1)}
\Psi(x;a,\tau;\ep_1)\,\psi_v(t;a,\tau;\ep_1)\big(1+\CO(\ep_2)\big)\,,
\end{equation}
where $\mathcal{Y}(a;\tau;\ep_1)$ and $\Psi(x;a,\tau;\ep_1)$ coincide with the 
functions appearing  in \rf{NSlim}, and
$\psi_v(t)=\psi_v(t;a,\tau;\ep_1)$ generate a basis for the space of solutions of the differential equation 
$(\ep_1^2\partial_t^2-u(t))\psi_v(t)=0$
where $u=u(t)$ represents 
the oper connection %corresponding to a point $a$ in ${\rm Op}_{\fsl_2}(C_\tau)$ 
as $\ep_1\partial_t+\big(\begin{smallmatrix} 0& u \\ 1 & 0\end{smallmatrix}\big)$.
The index $v=+,-$ labels a basis for the space of solutions. 

The fact that $\psi_v(t)$ factors off is related to the Hecke eigenvalue property, 
as has been pointed out in  \cite{T10}. As before  we may interpret 
$\mathcal{Z}_v^*$ as overlap $\langle \de_x^{\sst (2)},\xi_a^{}\rangle$,
where $\xi_a\in\CH^{\sst (2)}_a$. If \rf{NSlim+} holds for all $x$, it implies a 
similar factorisation for all $\xi_a\in\CH^{\sst (2)}_a$. 
The  factor $\psi_v(t)$ does furthermore not depend on the choice of
vector $\xi_a\in\CH^{\sst (2)}_a$ used  to represent 
the partition function as $\langle \de_x^{\sst (2)},\xi_a^{}\rangle$.
The monodromy  generated by analytic continuation of $\psi_v(t)$
defines an oper local system $\chi$. Being independent of  $\xi_a\in\CH^{\sst (2)}_a$, 
one may attribute this modification to a modification of the space $\CH^{\sst (2)}_a$ 
itself, satisfying the Hecke eigenvalue property.

%{\crd Discuss relation to Hecke eigenvalue property.}

\subsection{Hecke eigenvalue property from the $tt^*$-equations}\label{Hecke-4d}

Our goal in this section will be to  elaborate on the (4d+2d)-description of the Hecke functors.
Due to the fact that the $\Omega$-deformation localises fluctuations to the 
fixed points of the circle actions, one may think of the limit $\ep_2\ra 0$ as a
decompactification of the  hemisphere $B^4_{\ep_1\ep_2}$ into 
$B^2_{\ep_1}\times\BR^2$, where $B^2_{\ep_1}$ is a two-dimensional 
hemisphere. The surface operators of interest 
will be supported on $B^2_{\ep_1}$.

The goal is to offer an interpretation for the behaviour \rf{NSlim+} in four-dimensional terms. Recall that the 
partition function $\mathcal{Z}_v(a;x,\tau,t;\ep_1,\ep_2)$ can be interpreted as a partition function of the gauge
theory on  a 4d hemisphere $B^4_{\ep_1\ep_2}$ with $\Omega$-deformation, 
defined using boundary conditions projecting 
the state created by the gauge theory path integral on $B^4_{\ep_1\ep_2}$ to an eigenstate of the 
Wilson-loops \cite{NW,TV13}. The  variables $a$ parameterise the Wilson loop eigenvalues, 
and the collection of UV gauge couplings of the 4d SUSY gauge theory is  denoted by $\tau$.
%The 4d definition  of the 
%codimension-2 surface operator does not involve any new degrees of freedom on the surface, it was defined with the help of a boundary condition imposed on the bulk gauge fields \cite{AT}. 
The variables $x$ parameterise the boundary conditions defining
a codimension-2 surface operator \cite{FGT}, while the variable $t$ represents the 
FI parameter $t$ of the GLSM representing the additional surface operator of co-dimension four \cite{AGGTV}.

\subsubsection{Factorisation}

The degrees of freedom on the surface operator can not contribute to the divergence of the 
free energy when $\ep_2\ra 0$ as the two-ellipsoid appearing in this limit ends up having finite volume. 
We therefore expect that $\CF=\log\CZ$ behaves as
\begin{equation}
\mathcal{F}\sim-\frac{1}{\ep_2}\mathcal{Y}^{\rm 4d}(a;\tau;\ep_1)+\mathcal{W}^{\rm 2d}(a,\tau;x,t;\ep_1)+\dots\,,
\end{equation}
where $\mathcal{W}^{\rm 4d}(a;\tau;\ep_1)$ is independent of the 2d couplings $x$ and $t$. Comparison with 
\rf{NSlim} gives
\[
\mathcal{W}^{\rm 2d}(a,\tau;x,t;\ep_1)=\mathcal{W}^{\rm\sst M2}(a,\tau;t;\ep_1)
+\mathcal{W}^{\rm\sst M5'}(a,\tau;x;\ep_1)\,, 
\]
where $\mathcal{W}^{\rm\sst M2}(a,\tau;t;\ep_1)=\log\psi_v(t;a,\tau;\ep_1)$ and 
$\mathcal{W}^{\rm\sst M5'}(a,\tau;x;\ep_1)=\log\Psi(x;a,\tau;\ep_1)$. 
This decoupling can probably be understood as a consequence of the 
fact that the boundary condition defining a codimension 2 surface operator only involves the bulk fields
and not the degrees of freedom used to describe the codimension 4 surface operator.

It is then natural to interpret $\mathcal{Z}^{\rm 2d}_v(a,\tau;t;\ep_1):=\psi_v(t)$ as the partition function of the 
GLSM on the lower 2d hemisphere $B^2_{\ep_1}$ with $\Omega$-deformation, 
coupled to the class $\CS$ theory in the bulk, and subject to boundary 
conditions denoted as $\mathfrak{b}_v$. Examples of 
such partition functions have recently been studied in  \cite{FKNO}, where a close relation
was found to the hemisphere partition functions studied in \cite{HO,HR}. It was observed that 
the dependence of $\mathcal{Z}^{\rm 2d}_v(a,\tau;t;\ep_1)$ on the FI parameter $t$
is holomorphic. The partition function is furthermore related to an overlap in the topological twisted version 
of the GLSM as \cite{HO,HR}
\begin{equation}\label{Zoverlap}
\mathcal{Z}^{\rm 2d}_v(a,\tau;t;\ep_1)\,=\,\langle\,\mathfrak{b}_v\,|\,0\,\rangle\,,
\end{equation}
where $|\,0\,\rangle$ is the state created 
by the path integral on the hemisphere with no operator insertion, and 
$\langle\,\mathfrak{b}_v\,|$ is the boundary state associated to the boundary 
condition $\mathfrak{b}_v$. 
%The arguments used to derive \rf{Zoverlap} in \cite{HO,HR}
%generalise  similar arguments previously used in \cite{GL} 
% for the case of the sphere partition function.

\subsubsection{$tt^*$-equations}

The overlap $\langle\,\mathfrak{b}_v\,|\,0\,\rangle$ has a natural generalisation obtained by replacing 
$|\,0\,\rangle$ by an other supersymmetric ground state $|\,1\,\rangle$ created from $|\,0\,\rangle$
by means of the chiral ring generator $v$, $|\,1\,\rangle= v|\,0\,\rangle$. 
It has been observed in 
\cite{HIV} that 
$\Pi_i^v:=\langle\,\mathfrak{b}_v\,|\,i\,\rangle$ represents a horizontal section of the $tt^*$-connection of 
Cecotti and Vafa \cite{CV}. 
In our case we  have a single coupling $t$ and a two-dimensional 
space of SUSY vacua.
The $tt^*$-connection therefore has the form 
of a flat $SL(2,\mathbb{C})$-connection $\nabla'=(\partial_t+\mathcal{A}_t)dt$, 
$\nabla''=(\bar\partial_{\bar{t}}+\mathcal{A}_{\bar{t}})d{\bar{t}}$ \cite{Ga09}.
%The basis generated by
%the chiral ring generators exhibits the holomorphic structure of the bundle of ground states, 
%Setting $\mathcal{A}_{\bar{t}}$ to zero by a gauge transformation makes $\mathcal{A}_t$ holomorphic.
It follows from 
the holomorphicity observed in  \cite{FKNO,HO,HR} that $\mathcal{A}_{\bar{t}}$ must be nilpotent 
in the basis $\Pi_i^v$ for the 
case at hand: This is necessary for the existence of a gauge transformation 
preserving $\Pi_0^v$ and relating the connection $\mathcal{A}$ to an oper. The horizontality condition 
$\nabla' \Pi^v=0=\nabla''\Pi^v$ then implies a second order differential equation for $\Pi_0^v$. 
The observations above give a physical realisation for the description of opers given in \cite{DFKMMN}.

It follows from the arguments presented in \cite{AGGTV} that the couplings $t$ can be identified with 
local coordinates on the Riemann surface $C_\tau$.  On physical grounds one would expect that the 
ordinary differential equation following from the horizontality condition should be non-singular 
on $C_\tau$. This would imply
that the $tt^*$-connection must be gauge equivalent to an oper connection. 
The results of \cite{FKNO,HO,HR} provide support for this claim.
%In Appendix \ref{hemiapp}
%we discuss the relation between this claim
%and the known results on hemisphere partition functions in more detail.

\subsubsection{Fixing the monodromies}

A weakly coupled Lagrangian description of the 4d theories of class $\mathcal{S}$ exists when the complex structure 
of $C_\tau$ is near a boundary of the moduli space of Riemann surfaces where $C_\tau$ can be represented 
as a collection of pairs of pants connected by long tubes $T_1,\dots,T_{3g-3+n}$.  
The coupled 4d-2d-system will be weakly coupled 
if $t$ is located on a particular tube $T_r$. The Lagrangian  description involves a
coupling of the GLSM to the 4d bulk theory of the form
$ta_r/\ep_1$, where $a_r$ is the restriction of the scalar field in the vector multiplet associated to $T_r$
to the support of the codimension-4 surface operator \cite{AGGTV}. The monodromy of the partition function
$\mathcal{Z}^{\rm 2d}_v(a;\tau;t;\ep_1)$ corresponding to $t\ra t+2\pi$ is therefore diagonal, and can be represented
by multiplication with $e^{2\pi va_r/\ep_1}$. This fixes half of the monodromy of the $tt^*$-connection which 
is enough to fix it completely,  being gauge equivalent to an oper connection. 

The  $tt^*$-connection thereby gets identified with the oper local system $\rho$ appearing on one
side of the geometric Langlands correspondence.  Taking the quotient of the algebra 
of global differential operators on ${\rm Bun}_G$ by the ideal generated from
$D_r-E_r$, $r=1,\dots,3g-3+n$ defines the corresponding D-module $\De_{\rho}$ on the other
side of the 
Langlands correspondence \cite{Fr07}.
% The D-module $\De_{\rho}$ defines a  
% flat connection on 
%a vector bundle over the complement of a singular locus in ${\rm Bun}_G$.
%The solutions $\Psi(a;\tau;x)$ of the eigenvalue equations 
%$\mathsf{H}_r\Psi=E_r\Psi$ represent horizontal sections of this connection.
The factorisation \rf{NSlim+} %of the chiral partition functions 
can  be 
seen as a manifestation of the Hecke eigenvalue property 
on the level of the solutions of differential equations associated to the D-module $\De_{\rho}$.

\section{Sigma model interpretation?}
 
The discussion in the previous section has revealed close connections
between generalisations of the AGT-correspondence in the presence of 
surface operators and the Beilinson-Drinfeld approach to the geometric 
Langlands correspondence. We had identified the geometric Langlands 
correspondence as the limit $\ep_2\ra 0$ of a description for the spaces of 
conformal blocks in terms of the quantised character varieties.
In this final section we are going to return to the question how the
emerging picture 
is related to the Kapustin-Witten approach to the geometric 
Langlands correspondence.  In Section \ref{comp-I} we will raise the
question how to describe the limit $\ep_2\ra 0$ in terms of the $2d$ sigma model. 
Sections \ref{Cstaraction} and \ref{SecNSlim} outline a speculative answer to this question. 
In the rest of this Section we will propose a more explicit comparison between the 
approaches of Beilinson-Drinfeld and Kapustin-Witten based on similar arguments.

\subsection{Comparison I: Relations between AGT- and geometric Langlands correspondences}\label{comp-I}

We had argued that the $\CD$-modules appearing in this 
context can be characterised in terms of the spaces of solutions
to the corresponding differential equations, physically realised as
partition functions. In the context of the AGT-correspondence with 
surface operators we had argued that the spaces  
$\CH_{x}^{\sst(2)}:=
{\rm Hom}_{\CM_H(G)}^{}(\mathsf{B}_{\rm cc}^{},\mathsf{L}_{x}^{\sst(2)})$
can be identified with the fibers of affine algebra conformal blocks over 
the bundles $\CE_x$. In CFT we can 
view the chiral partition functions $\CZ(a;x,\tau;k)$ solving the 
KZB equations as representatives of elements $\vf_a$ of subspaces $\mathcal{CB}^\si_a(C,\hfg_k,\CE_x)$
of the spaces $\mathcal{CB}(C,\hfg_k,\CE_x)$ conformal blocks having fixed eigenvalues parameterised by
$a$ under a maximal commuting subset of the Verlinde line operators.

 Modular invariance
of the open sigma model TQFT suggests that we can identify 
$\mathcal{CB}^\si_a(C,\hfg_k)\simeq 
{\rm Hom}_{\CM_H(G)}^{}(\mathsf{B}_{\rm cc}^{},\mathsf{L}_{\si,a}^{\sst(1)})$. 
This identification can be supported  by observing that the boundary condition
$\mathsf{L}_{\si,a}^{\sst(1)}$ fixes the values of some zero modes in the sigma model
on the strip, which should lead to an eigenvalue property w.r.t. to the subalgebra of 
$\mathcal{A}_{B}^{\hbar}$ generated by the corresponding quantised trace functions, associated
to the curves defining the pants decomposition $\si$.

In the last section we 
have discussed how the subspaces $\mathcal{CB}_a^\si(C,\hfg_k,\CE_x)$ get related to 
the Hecke-eigen $\CD$-modules appearing the 
the geometric Langlands correspondence in the limit $\ep_2\ra 0$, with 
local systems representing the ``eigenvalue'' being represented by families of 
opers parameterised by the variables $a$.
What is not clear, however, is how our observations concerning the appearance of the Hecke
eigenvalue property in the limit $\ep_2\ra 0$ can be understood from the perspective
of the sigma model. There appears to be an immediate obstacle: 
The sigma model considered in \cite{KW} in the context of the 
geometric Langlands correspondence are A-models with 
symplectic structure $\omega_K$, and the cc-brane has 
Chan-Paton curvature $\omega_J$. A different choice 
appears in the Nekrasov-Witten approach to the AGT-correspondence,
where the A-model with 
symplectic structure $\frac{\ep_2}{\ep_1}\omega_I$ is used. 
There is no obvious parameter allowing us to move
continuously between these two cases. 

In view of the fact that within CFT one can obtain the  
geometric Langlands correspondence in 
the critical level limit it seems very natural to ask if this 
can be understood within the $2d$ sigma model with target $\CM_H(G)$.
Starting from Subsection \ref{Cstaraction} we'll speculate about
a possible way to see this.

\subsection{Hyperk\"ahler rotation}\label{Cstaraction}

It is tempting to modify the 2d TQFT set-up of Nekrasov and Witten using the 
hyperk\"ahler$^{(g)}$ rotations coming from the circle action $\vf\ra e^{i\theta}\vf$ on Hitchin's 
moduli spaces. It is 
shown in \cite[Section 9]{Hi87} that this circle action can be complexified into a 
$\BC^*$-action relating all complex and symplectic structures in the family $(I_\zeta,\Omega_\zeta)$
apart from $\zeta=0,\infty$.
The hyperk\"ahler rotations allow us to identify
the key geometric structure on $\CM_H(G)$ for $\zeta\neq 0,\infty$
in the sense that the rotations can be represented
by diffeomorphisms of $\CM_H(G)$ \cite[Proposition (9.1)]{Hi87}.
This amounts to reparametrizations of 
the sigma model fields, which may not affect the 
physical content. 

Invariance of the sigma model path integral under field redefinitions 
may be a subtle issue. The circle action $\vf\ra e^{i\theta}\vf$ should be easy to 
understand, but it is not clear to us if the complexification to a $\BC^*$-action
is easy to understand as a symmetry of the sigma model TQFT in a suitable sense.
What we'd like to 
point out is that validity of a certain form of invariance of the topological
sigma models with target $\CM_H(G)$ under the $\BC^*$-action
would give us a simple way to interpolate between the sigma models considered
in \cite{KW} and \cite{NW}, respectively. In this way one could get an 
attractive explanation of the 
observation made in the previous section.

We are interested in open A-models on surfaces with boundaries with boundary conditions
being either of the canonical coisotropic or Lagrangian type. We may use the family of 
holomorphic symplectic forms $\Omega_{\zeta}$ to define a family of 
A-models with symplectic structure $\mathrm{Im}(\Omega_\zeta)$ having
coisotropic  branes with Chan-Paton curvature $\mathrm{Re}(\Omega_\zeta)$. 
Submanifolds of $\CM_H(G)$  which are Lagrangian with respect to $\Omega_\zeta$
define branes in this family of A-models.
If an appropriate form of $\BC^*$-invariance of the sigma model TQFT holds, one would expect that the 
partition functions $\CZ_\zeta$ defined in this family of A-models are in fact 
$\zeta$-independent.  

One should note, however, that all Lagrangian submanifolds we use to define
boundary conditions will have to be varied consistently to keep $\CZ_{\zeta}$ unchanged.
The submanifolds of our interest are the orbits of the complex Fenchel-Nielsen$^{(g)}$
twist flows in the character variety, considered as submanifolds of $\CM_H(G)$ via
NAH-correspondence and holonomy map. Considering a fixed orbit 
in the character variety, one gets a one-parameter family of submanifolds of $\CM_H(G)$ 
upon varying the hyperk\"ahler parameter $\zeta$.

Given we have invariance under the $\BC^*$-action, we could combine variations of the
coupling parameter $\ep_1/\ep_2$ with a suitable hyperk\"ahler rotation in such a way that 
we obtain a one-parameter family of topological A-models 
with target $\CM_H(G)$ that interpolates between the ones considered in
\cite{NW} and \cite{KW}, respectively. This could be done by setting $\zeta=\ep_2\xi$. 
The expression for 
$\hat{\omega}_\zeta=\frac{\ep_2}{\ep_1}\mathrm{Im}(\Omega_{\zeta})$
reduces in the limit
$\ep_2\ra 0$ to $\hat{\omega}_\zeta=
\frac{1}{2\ep_1|\xi|^2}
(\mathrm{Re}(\zeta)\omega_K-\mathrm{Im}(\zeta)\omega_J)$, reproducing 
for real $\xi$ the 
symplectic form used to define the A-model studied in \cite{KW}.

\subsection{Nekrasov-Shatashvili limit of the branes $\mathsf{L}_{\si,a}^{\sst (1)}$} \label{SecNSlim}

In order to see that the resulting scenario may indeed resolve the puzzle stated above,
let us first note that the (A,B,A) branes 
$\mathsf{L}_{\si,a}^{\sst (1)}$ considered in Section \rf{branes1} can easily be
generalised into $(B,A,A)_\zeta$-branes $\mathsf{L}_{\si,a}^{\sst (1)\zeta}$ by 
using the $\zeta$-dependent NAH-correspondence in their definition.
Choosing $\zeta$ in an $\hbar$-dependent way, as suggested above, would turn them
into branes $\mathsf{L}_{\si,a}^{\sst (1)\hbar}$ in the family of A-models with 
hyperk\"ahler parameter $\hbar=\ep_1/\ep_2$.

We will observe that something interesting happens in the 
Nekrasov-Shatashvili of the family of branes $\mathsf{L}_{\si,a}^{\sst (1)\hbar}$:
The  
branes  $\mathsf{L}_{\si,a}^{\sst(1)\hbar}$ will have a well-defined limit $\hbar\ra \infty$ if
the parameters $a$ are scaled in this limit as 
${a}_r=\frac{\ep_1}{\ep_2}\,\check{a}_r$ with $\check{a}_r$  finite.
In order to understand the limit $\ep_2\ra 0$ one mainly needs to study the WKB approximation
for the holonomy of $\pa_u+\mathcal{A}_u$ where $\mathcal{A}_u=A_u+\frac{1}{\ep_2}\vf$. 
By gauge transformations one can always locally reach the form 
$\mathcal{A}_u=\big(\begin{smallmatrix} 0 & u_{\ep_2} \\ 1 & 0\end{smallmatrix}\big)$. 
The function  $u_{\ep_2}(t)$ appearing in the upper right matrix element of $\mathcal{A}_u$ 
will in general only be meromorphic in $u$, but the residues are of order $\ep_2$.
We note that $A_u$ contributes only in subleading orders of the 
expansion. 
It follows easily from these facts that $\vartheta_{\ep_2}(t)=\frac{1}{2}{\rm tr}(\vf^2)+\CO(\ep_2)$.
To leading order in $\ep_2$ one may  therefore represent the solutions $s$ to
$(\pa_u+\mathcal{A}_u)s_{\ep_2}=0$ in the form
\begin{equation}
s_{\ep_2}(t)\,\sim\, e^{\pm\frac{1}{\ep_2}\int_{\CC_t} \la}\,,
\end{equation}
where $\CC_t$ is a path on the spectral curve $\Sigma$ ending at a lift $\hat{t}\in\Sigma$ of $t\in \C$,
and $\la$ is the canonical differential on $\Sigma$.
The conclusion is that the rescaled Fenchel-Nielsen length coordinates $\check{a}_r$
behave in the limit $\ep_2\ra 0$ as
$
\check{a}_r=\frac{1}{\ep_1}\mathsf{a}_r+\CO(\ep_2),
$
where $\mathsf{a}_r$ are the periods of the  differential $\la$ 
defined along a Lagrangian subspace of a canonical homology basis
determined by $\si$.

One may in this sense view the ``quantum periods'' $\check{a}_r$ as deformations of the angle variables 
$\mathsf{a}_r$
for the Hitchin integrable system. It is interesting to note that the dependence on $\si$
disappears in the limit ${\ep}\ra 0$: Even if the definition of the coordinates $\mathsf{a}_r$
carries a residual dependence on $\si$, this is not the case for the fiber of Hitchin's fibration
determined by the values of the $\mathsf{a}_r$.
In this way we are led to the conclusion  that the scaling
limits of the branes $\mathsf{L}_{\si,a}^{\sst(1)\hbar}$ get represented by the (B,A,A)-branes
supported on the fibers $\mathsf{F}_{\sst (u_{\mathsf{a}},0)}$
of Hitchin's (first) fibration with {trivial}
Chan-Paton bundle over a point $u_\mathsf{a}$ on the Hitchin base $\CB$  determined by the 
coordinates $\mathsf{a}_r$.
%{\color{red} $\leftarrow$ more explanations why ``trivial"?}

Let us finally note that one could discuss 
the branes $\mathsf{L}_{x}^{\sst(2)\hbar}$ in a similar way. It is easy to see that one 
obtains the branes 
$\mathsf{L}_{x}^{\sst(2)}$ supported on fibers of Hitchin's second 
fibration when ${\hbar}\ra 0$.

\subsection{Comparison II: Relation between Kapustin-Witten approach and CFT?}\label{compII}

Any comparison between the approaches of Beilinson-Drinfeld and Kapustin-Witten
will need to address the
following point. Hecke-eigenbranes are described in \cite{KW} as 
skyscraper sheaves on $\CM_H(G)$. However, in order to use SYZ mirror
symmetry on the Hitchin fibration, Kapustin and Witten use the representation
of $\CM_H(G)$ and $\CM_H({}^L G)$ adapted to the complex structure $I$, whereas the Beilinson-Drinfeld 
approach considers ${}^LG$-local systems $(\CE,\nabla')$
on one side of the correspondence. In order 
to relate the two, one needs to use the NAH-correspondence. 
If $\mu$ represents a point of $\CM_H(G)$, represented as a torus 
fibration,  let $\check{\mathsf{F}}_{\mu}$ be the corresponding skyscraper 
sheaf in the B-model, and $\mathsf{F}_{\mu}$ be the Hitchin fiber in 
the SYZ-dual A-model. In order to formulate a conjectural 
relation between the two approaches 
we need to use the (inverse of the) NAH-correspondence to find the point $\mu(\chi)\in\CM_H(G)$
associated to the local system $\chi$.

A natural guess for a possible relation between
the approaches of Beilinson-Drinfeld and Kapustin-Witten
could then be the validity of the isomorphism of $\CD$-modules 
\begin{equation} \label{BD-KW}
\begin{aligned}
{\rm Hom}_{\CM_H(G)}^{}(\mathsf{B}_{\rm cc},\mathsf{F}_{\mu(\chi)}) 
& \;\simeq\; \mathcal{CB}(C,\hfg_{-h^{\vee}},\CE)\;\;\text{for}\;\;\CE\in\mathrm{Bun}_G^{\rm vs}\\
& \;\simeq\; \big[\,
\text{Fiber of}\;\, \De_\chi\;\,\text{over}\;\,\CE\in\mathrm{Bun}_G^{\rm vs}\,\big]
\end{aligned}
\end{equation}
where $\mathrm{Bun}_G^{\rm vs}$ is the space of
``very stable'' bundles $\CE$, bundles that do not admit a nilpotent Higgs field,
and $\De_\chi$ is the $\CD$-module represented by % the conformal blocks in 
$\mathcal{CB}(C,\hfg_{-h^{\vee}},\CE)$.
The right hand side does not depend on the choice 
of $\CE$ due to the existence of a canonical flat connection  identifying fibers
associated to different $\CE\in\mathrm{Bun}_G^{\rm vs}$.
The dimension of $\mathcal{CB}(C,\hfg_{-h^{\vee}},\CE)$  may jump 
away from $\mathrm{Bun}_G^{\rm vs}$,  see \cite[Section 9.5]{Fr07} for a discussion.

At this point we may note another puzzle arising in the comparison of the 
approaches of Beilinson-Drinfeld and Kapustin-Witten. As noted above, 
and illustrated by the examples studied in \cite{T10}, there is 
a somewhat discontinuous behaviour of the $\CD$-modules appearing in 
the Beilinson-Drinfeld approach to the 
geometric Langlands correspondence away from the sub-variety of opers within
$\CM_{dR}({}^LG)$, described by the appearance of a number of additional 
degenerate representations in the representation of the $\CD$-modules as 
conformal blocks. 
No such discontinuous behaviour is seen in the approach of Kapustin and Witten.

\subsection{Conformal limit}

We would also like to suggest a way which might lead to an answer for the 
questions raised in Subsection \ref{compII} above. It is based on the observation 
made in \cite{Ga14,DFKMMN} that the NAH correspondence may simplify drastically
in the conformal limit where the parameter $R$ introduced into the NAH correspondence 
by scaling $\vf\ra R\vf$ is sent to zero together with the hyperk\"ahler
parameter $\zeta$ such that $\zeta/R$ stays finite.
As a preparation we'd here like to discuss the possible 
relevance of this limit. The issues are similar, but non-identical to the ones
concerning the possible relevance of hyperk\"ahler rotations discussed
in Section \ref{Cstaraction}, motivating a separate discussion.

Replacing $\vf\ra R\vf$ defines a one-parameter family of deformations of 
the NAH correspondence, leading to an apparent modification of the hyperk\"ahler metric
on $\CM_H(G)$. However, the parameter $R$ is inessential for the geometry of Hitchin's moduli spaces
in the sense that hyperk\"ahler metrics associated to different values of  $R$ are related
by diffeomorphisms. 
One may define one-parameter families of sigma model 
actions $S_R$ using the hyperk\"ahler metrics on $\CM_H(G)$ 
obtained from the modified NAH correspondence. However,
being constructed out of 
hyperk\"ahler metrics related by
diffeomorphisms, one may be tempted to 
identify two actions $S_{R_1}$ and $S_{R_2}$ differing only 
in the choice of the parameter $R$ 
as physically equivalent Lagrangian representations for the {\it same} sigma model QFT. 
In suitable coordinates like  $u=\frac{1}{2}\mathrm{tr}(\vf^2)$ one finds that
the relevant diffeomorphism is realised as a simple scaling $u\ra ur^{2}$. 
The corresponding field redefinition should indeed lead to a rescaling of 
partition functions by inessential overall factors only.

If the partition functions $Z_R(u)$ defined using the actions $S_R$ 
depend on a boundary parameter $u$ associated to 
a coordinate for $\CM_H(G)$ scaling under $\vf\ra R\vf$ as $u\ra ur^{2}$,  
we find that the $R$-independence of the sigma model metric modulo diffeomorphisms
is expressed in the fact that $Z_{r R}(u)\propto Z_R(ur^{2})$, 
again possibly up to inessential overall factors.

A partition function $Z'(\chi)$ in the boundary B-model with a boundary 
condition defined by a point $\chi$ on $\CM_{dR}({}^LG)$
may be represented in terms of a partition function $Z_R''(\mu_{R,\zeta}(\chi))$ depending on 
a point $\mu_{R,\zeta}(\chi) \in \CM_{H}({}^LG)$ since the $R$-dependence resulting from the 
modification of the NAH correspondence is compensated by the corresponding 
modification of the sigma model action. This observation may be
useful if there is a limit where $\mu_{R,\zeta}(\chi)$ simplifies considerably. 
We will see that such a limit is the so-called conformal limit $R\ra 0$,
$\zeta\ra 0$ keeping $\zeta/R$ finite. This will lead to an interesting
reformulation of the proposed relation \rf{BD-KW}, as we shall now discuss.

\subsection{Relation between Kapustin-Witten approach and CFT, more explicitly}

We will mostly restrict attention to the case where the local systems $\chi$ are opers 
$\rho_u=\big(\CE_o,\big(\pa_t+\big(\begin{smallmatrix} 0& u\\ 1 & 0\end{smallmatrix}\big)\big)dt\big)$, 
playing a basic role 
in the approach of Beilinson and Drinfeld. It is interesting to note that the conformal limit of the
NAH correspondence becomes particularly simple in this case \cite{DFKMMN}, relating opers 
$\rho_u$ 
to Higgs pairs
of a very particular form. For ${}^L G=\mathrm{PSL}(2,\BC)$ one finds Higgs pairs $(\mathcal{E},\vf)$ of the form 
$(K^{1/2}\oplus K^{-1/2},(\begin{smallmatrix} 0& u \\ 1 &0\end{smallmatrix})dz)$, 
where $u$ is the quadratic differential representing $\rho_u$ in the Fuchsian uniformisation of $C$.
Passing to the description of $\CM_H({}^L G)$ as a torus fibration represented by 
pairs $(\Sigma_u,\CL)$, where $\Sigma_u$ is the spectral curve
associated to a point $u\in\CB$ of the Hitchin base, and $\CL$ is a line bundle on
$\Sigma_u$, one gets the line bundle $\CL_0=\pi^*(K^{1/2})$, with $\pi: \Sigma_u\ra C$
being the covering projection. This line bundle represents a canonical 
``origin'' of the Jacobian/Prym  parametrising the choices of $\CL$ \cite{W15}. 

Let $\check{\mathsf{F}}_{\sst(u,0)}$ be the 
skyscraper sheaf on $\CM_{H}({}^LG)$ supported at the 
point $(\Sigma_u,\CL_0)$ and let
$\mathsf{F}_{\sst (u,0)}$ be the fiber of the Hitchin fibration which is SYZ dual to 
$\check{\mathsf{F}}_{\sst(u,0)}$. % supported at the point $(u,0)$ of the Hitchin fibration.
The possible relation \rf{BD-KW}
between the approaches of Beilinson-Drinfeld and Kapustin-Witten may then be
formulated more explicitly as 
\begin{equation} \label{BD-KW-conf}
{\rm Hom}_{\CM_H(G)}^{}(\mathsf{B}_{\rm cc},\mathsf{F}_{\sst (u,0)})
\;\simeq\;
\big[\,
\text{Fiber of}\;\, \De_{\rho_u}\;\,\text{over}\;\,\CE\in\mathrm{Bun}_G^{\rm vs}\,\big]\,.
\end{equation}
We remark that the image of generic points $(u,0)\in\CM_{H}({}^LG)$ under 
the NAH correspondence will be represented by an oper 
if and only if the map from $\CM_{H}({}^LG)$ to $\CM_{dR}({}^LG)$
is defined using the conformal limit of the non-abelian Hodge correspondence.\footnote{The 
direction ``if'' was shown in  \cite{DFKMMN}. The image of points $(u,0)\in\CM_{H}({}^LG)$ under NAH
consists of connections with real holonomy, intersecting the variety of opers only 
discretely. We thank A. Neitzke for this remark.} This, and the relevance of this limit in Section \ref{Hecke-4d}, indicate that this limit is well-suited for formulating the 
relation between the 
approaches of Beilinson-Drinfeld and Kapustin-Witten. 

Concerning the generalisation of \rf{BD-KW-conf} to more general local systems $\chi$
we conjecture  that there exist natural stratifications 
of $\CM_{dR}({}^LG)$ and $\CM_{H}({}^LG)$, having strata related to each other by
the conformal limit of the NAH correspondence. This would allow us to extend the 
relation  \rf{BD-KW-conf} to generic irreducible local systems, linking the discontinuous 
behaviour of the $\CD$-modules appearing in the geometric Langlands correspondence to 
the passage from one stratum to another. We plan to return to this 
point in a forthcoming publication.

\subsection{Outlook}

We will elsewhere discuss available evidence for the  existence of 
relations of the  form 
\begin{equation}\label{spectral}
\CH_{x}^{\sst(2)}\simeq \int^{\oplus} d\mu_\si(a)\;\CH_{\si,a}^{\sst(1)}\,,
\end{equation}
and for the existence of linear relations between
the spaces 
$\CH_{\si,a}^{\sst(1)}$ associated to  different pants decompositions $\si$.
This
restores a  weaker
version of $\si$-independence within the story associated to nonzero $\epsilon_2$.
%\end{rem}

The geometric Langlands correspondence is sometimes presented as an analog
of the spectral decompositions of spaces of automorphic forms appearing in the 
classical Langlands program. We view the contents of this note as hints that
it may not be outrageous to dream of a variant of the 
geometric Langlands program extending it by transcendental and analytic
aspects. The transcendental aspects involve the partition functions
representing solutions to the systems of differential equations defined by the $\CD$-modules,
and the analytic aspects concern spectral decompositions as proposed in \rf{spectral}.
Identifying the partition functions as analogs of the automorphic forms
would  strengthen the analogies to the 
classical Langlands program even further. The partition functions represent the 
bridge between the algebraic structures of $\CM_H(G)$ associated 
to the representation as moduli space $\CM_{dR}(G)$ of local systems, and as 
character variety $\CM_B(G)$, respectively. In this way one may expect to get 
a larger picture unifying topological and complex structure dependent 
aspects of the geometric Langlands program.

We plan to discuss these matters, 
the interpretation as ``quantum geometric Langlands duality'',  and the relation
to another incarnation of 
Langlands duality patterns referred to as modular duality in \cite{T10} 
in forthcoming publications.

Let us finally note that recent progress on the geometric Langlands program from the 
gauge theory side has been described in \cite{Ga16a,Ga16b}. It should be interesting
to analyse the relations to our work.

{\bf Acknowledgements:}  
A.B would like to thank D. Ben-Zvi, M. Mulase, A. Neitzke and R.
Wentworth for discussions and the organizers of String-Math 2016 for
putting together a stimulating conference. A.B would also like to thank
L'Institut Henri Poincar\'e (Paris) and ICTS-TIFR (Bangalore) for
hospitality during visits when this work was in progress.

J.T. would like to thank the organizers of String-Math 2016 for setting up an inspiring conference and the 
opportunity to present
this work,  and 
M. Mulase, A. Neitzke for discussions. Special
thanks to E. Frenkel for various discussions, for
communicating the content of his unpublished work \cite{Fr10},
and for critical remarks on the draft.

This work was supported by the Deutsche Forschungsgemeinschaft (DFG) through the 
collaborative Research Centre SFB 676 ``Particles, Strings and the Early Universe'', project 
A10.
\appendix

\section{Hitchin's moduli spaces} \label{Gloss}

We assume that $G=SL(2)$, and that $C$ is a Riemann surface with genus $g$ 
and $n$ punctures.

{\bf  Hitchin  moduli space $\CM_H(G)$} \cite{Hi87}. 
Moduli space of pairs $(\CE,\vf)$, where $\CE=(E,\bar{\partial}_{\CE})$ is a holomorphic structure on 
a smooth vector bundle $E$, and $\vf\in H^0(C,\mathrm{End}(\CE)\ot K)$. The moduli space of such 
pairs modulo natural gauge transformations is denoted by $\CM_H(G)$.

{\bf  Hitchin's integrable system} \cite{Hi87}.
Given $(\mathcal{E},\vf)$ one constructs the spectral curve 
$\Sigma=\{(u,v);v^2=\frac{1}{2}\mathrm{tr}(\vf^2)\}\subset T^*C$, 
and the line bundle $\CL$ representing the 
cokernel of  $\vf-v$. 
One may reconstruct $(\mathcal{E},\vf)$  from $(\Sigma, \mathcal{L})$
as
$\mathcal{E}=\pi_*(\mathcal{L})$ and 
$\vf=\pi_*(v)$.
This describes  $\CM_{H}(G,C)$ as a torus fibration over the base
$\CB\simeq H^0(C,K^2)$, 
with fibres representing the choices of $\CL$ 
identified with the Jacobian of $\Sigma$ if
$G=GL(2)$, and with the Prym variety if $G=SL(2)$. Natural coordinates for the base $\CB$ are
provided by Hitchin's Hamiltonians, defined by expanding 
$\frac{1}{2}\mathrm{tr}(\vf^2)=\sum_{r=1}^{3g-3+n}\vartheta_rH_r$, with 
$\{\vartheta_r, r=1,\dots,3g-3+n\}$ being a basis for $H^0(C,K^2)$.

{\bf  Local systems}. Pairs  $(\CE,\nabla_\ep')$, where $\CE$ is a holomorphic vector 
bundle as above, and $\nabla_\ep'$ is a holomorphic $\ep$-connection, 
satisfying $\nabla_\ep' (fs)=\ep (\pa f)s+f\nabla_\ep's$ for 
functions $f$ and smooth sections $s$ of $E$. The moduli space of such 
pairs is denoted $\CM_{dR}(G)$.
Local systems are here often identified with the corresponding {\it flat bundles}, systems
of local trivialisations with constant transitions functions, or the representations 
of the fundamental group (modulo conjugation) obtained as holonomy of $(\CF,\nabla_\ep')$,
leading to the isomorphism between $\CM_{dR}(G)$ and the

{\bf Character variety $\CM_B(G)$}: The space of representations of $\pi_1(C)$ into $G$,
modulo overall conjugation, as algebraic variety described as a GIT quotient $\BC[\mathrm{Hom}(\pi_1(C),G]^G$.

{\bf  Opers}. Special local systems, where $\CE=\CE_{\rm op}$, the unique extension 
$0\ra K^{1/2}\ra\CE_{\rm op} \ra K^{-1/2}\ra 0$ allowing a holomorphic connection 
$\nabla_\ep'$ of the form $\nabla_\ep'=dz\big(\ep\pa_z+\big(
\begin{smallmatrix} 0 & u \\ 1 & 0\end{smallmatrix}\big)\big)$.

{\bf  Non-Abelian Hodge (NAH) correspondence} \cite{Hi87,Si}. Given a Higgs pair $(\CE,\vf)$, 
there exists 
a unique harmonic metric $h$ on $E$ satisfying 
%\begin{align}\label{selfdual}
$F_{\CE,h}+R^2[\vf,\vf^{\dagger_h}]=0$
%\end{align}
where $F_{\CE,h}$ is the curvature of the unique $h$-unitary connection $D_{\CE,h}$ having
$(0,1)$-part $\bar{\partial}_{\CE}$. % In the holomorphic gauge where $\bar{\partial}_{\CE}=\bar{\partial}$,
%we have $D_{\CE,h}=\partial+ h(\partial h^{-1}) +\bar\partial$, and \rf{selfdual} becomes
%\begin{align}\label{selfdual'}
%\bar\partial(h(\partial h^{-1}))+R^2\big[\vf,\bar{h}^{-1}\vf^{\dagger}\bar{h}\big]=0\,,
%\end{align}
%Equation \rf{selfdual} determines $h$ uniquely for given $(\CE,\vf)$.  
One may then form the corresponding two-parameter family of
flat connections
%\begin{align}
$\nabla_{\zeta,R}=\zeta^{-1}{R}\,\vf+D_{\CE,h}+R\zeta\,\vf^{\dagger_h}$.
%\end{align}
Decomposing $\nabla_{\zeta,R}$ into the $(1,0)$ and $(0,1)$-parts %$\nabla'$ and $\nabla''$ 
defines a pair $(\CF,\nabla_\ep')$ consisting
of $\CF=(E,\bar{\partial}_{\CF})$  and the $\ep$-connection 
$\nabla_\ep'=\ep\nabla'=\ep\pa_{\CE,h}+\vf$, with $\ep=\zeta/R$, holomorphic in
the complex structure defined by $\bar{\partial}_{\CF}$.

{\bf Hyperk\"ahler structure} \cite{Hi87}. There exists a $\mathbb{P}^1$ worth of complex 
structures $I_\zeta$ and holomorphic
symplectic structures $\Omega_\zeta$. The latter are defined as
$
\Omega_\zeta =\frac{1}{2}\int_C\;{\rm tr}(\de\mathcal{A}_\zeta\wedge\de\mathcal{A}_\zeta)
$.
A triplet of symplectic forms $(\omega_I,\omega_J,\omega_K)$ can be defined by expanding $\Omega_\zeta$ as
$
\Omega_\zeta=\frac{1}{2\zeta}(\omega_J+i\omega_K)+i\omega_I+\frac{1}{2}{\zeta}(\omega_J-i\omega_K).
$
The corresponding complex structures are
$
I_{\zeta}=\frac{1}{1+|\zeta|^2}((1-|\zeta|^2)I-i(\zeta-\bar{\zeta})J-(\zeta+\bar{\zeta})K).
$

{\bf Complex Fenchel-Nielsen coordinates} \cite{NRS}. Darboux coordinates 
for $\CM_{B}(G)$ associated to pants decompositions $\si$ of $C$ obtained
by cutting along closed curves $\ga_i$, $i=1,\dots,3g-3+n$.
The complex
length coordinates parameterise the trace functions $L_i=\mathrm{tr}(\rho(\ga_i))$
as $L_r=2\cosh(a_r/2)$. One may define canonically conjugate coordinates 
$\kappa_r$ such that the natural Poisson structure gets represented
as $\{a_r,\kappa_s\}=\de_{r,s}$, $\{a_r,a_s\}=0=\{\kappa_r,\kappa_s\}$.

\bibliographystyle{JHEP_TD}
\bibliography{AGT-GL}

\end{document}